\documentclass{IEEEtran}
\usepackage{graphicx}
\usepackage{cite}
\usepackage{mathtools}
\usepackage{blindtext}
\usepackage{amsmath,amssymb,amsfonts}
\usepackage{algorithm}
\usepackage{algpseudocode}
\usepackage{setspace}
\usepackage{subcaption}	
\usepackage{textcomp}
\usepackage{balance}
\usepackage{xcolor}
\usepackage{multirow}
\usepackage{stfloats}
\usepackage{amsmath}
\usepackage{array}
\usepackage{url}
\usepackage{soul}
\usepackage{tabularx}
\usepackage{bm}
\usepackage{pifont}%
\usepackage{hyperref}
\def\BibTeX{{\rm B\kern-.05em{\sc i\kern-.025em b}\kern-.08em T\kern-.1667em\lower.7ex\hbox{E}\kern-.125emX}}
    
\renewcommand{\baselinestretch}{0.94}

\begin{document}

\title{28~GHz Indoor and Outdoor Propagation Analysis at a Regional  Airport}

\author{
\IEEEauthorblockN{Kairui Du, Ozgur Ozdemir, Fatih Erden, and 
Ismail Guvenc}\\
\IEEEauthorblockA{Department of Electrical and Computer Engineering, North Carolina State University, Raleigh, NC}\\
Email: \{kdu, oozdemir, ferden, iguvenc\}@ncsu.edu
\thanks{This work was supported by NASA under the Award ID NNX17AJ94A. Authors would also like to thank the Johnston Regional Airport for allowing to carry out the experiments.}

}

\maketitle

\begin{abstract}
In the upcoming 5G communication, the millimeter wave (mmWave) technology will play an important role due to its large bandwidth and high data rate. However, mmWave frequencies have higher free-space path loss (FSPL) in line-of-sight (LOS) propagation compared to the currently used sub-6~GHz frequencies. What is more, in non-line-of-sight (NLOS) propagation, the attenuation of mmWave is larger compared to the lower frequencies, which can seriously degrade the performance. It is therefore necessary to investigate mmWave propagation characteristics for a given deployment scenario to understand coverage and rate performance for that environment. In this paper, we focus on 28~GHz wideband mmWave signal propagation characteristics at Johnston Regional Airport (JNX), a local airport near Raleigh, NC. To collect data, we use an NI PXI based channel sounder at 28~GHz for indoor, outdoor, and indoor-to-outdoor scenarios. Results on LOS propagation, reflection, penetration, signal coverage, and multi-path components~(MPCs) show a lower indoor FSPL, a richer scattering, and a better coverage compared to outdoor. We also observe high indoor-to-outdoor propagation losses.
\end{abstract}


\section{Introduction}
With the development of modern telecommunication, the use of wireless devices and applications that require higher data rates have increased tremendously in the recent decades. The sub-6 GHz frequency band is getting more congested by the rapid growth of users and it can not sustain the support of high data rates due to its limited channel bandwidth. The difficulty of supporting the demand of next generation communications at lower frequency bands motivated researchers to explore millimeter-wave~(mmWave) bands, 
which offer a substantially large amount of available bandwidth compared to sub-6 GHz frequencies. Owing to the millimeter-level wavelengths, large arrays of antennas can be used in small smart devices to achieve higher data throughput.


At sub-6 GHz frequencies, the wavelength is significantly larger compared to the mmWave bands which allows the signals to penetrate through obstacles in our surroundings such as walls, windows, doors, and foliage. On the contrary, the narrow wavelength of mmWave frequencies introduces propagation challenges, such as high free-space path-loss~(FSPL) and high attenuation while penetrating through different materials. It is hence necessary to fully study the propagation characteristics of mmWave signals in different environments. Fortunately, with the opening of mmWave spectrum by FCC~\cite{FCC_28G}, there has been a surge of channel measurements that report results at common mmWave frequency bands at 28, 39, 60, 73, and 91 GHz, which all help to analyze mmWave propagation. 

Reflection and penetration loss at 28 GHz in New York urban environment \cite{NYC} showed that the outdoor building materials are better reflectors than indoor materials, and that the penetration loss at larger distances are affected by the surrounding environment apart from distance and obstructions. Another study \cite{NYC2} reported that both outdoor non-line-of sight (NLOS) and line-of-sight (LOS) environments had rich multipath components at 28~GHz using steerable beam antennas. 
In \cite{hosseini2020attenuation}, authors focused on penetration loss of several typical building materials in three popular mmWave bands (28, 73, 91 GHz). As expected, higher penetration loss was observed as the frequency increases. Plywood and clear glass suffered higher attenuation (in dB/cm) compared to ceiling tile, drywall, and cinder blocks at all three frequency bands.
Another study \cite{xing2019indoor} focused on indoor reflection, penetration, scattering and path loss properties at both mmWave (28~GHz and 73~GHz) and sub-terahertz (140 GHz) frequencies. The authors found out that the reflection coefficient increases linearly as the incident angle increases. Reflection loss is lower at higher frequencies at a given incident angle while the penetration loss increases with frequency.
The authors in \cite{rappaport2015wideband} presented directional and omnidirectional path loss models, temporal and spatial channel models, and outage probabilities based on more than 15,000 measured power delay profiles (PDPs) at 28, 38, 60, and 73~GHz mmWave bands using wideband sliding correlator channel sounder and horn antennas. 

In our recent 28 GHz channel measurements \cite{library} at James B. Hunt Jr. Library of NC State University, path loss for the LOS scenarios was obtained to be very close to the free space path loss model. Models for power angular-delay profile (PADP) and large-scale path loss for both LOS and NLOS scenarios were developed based on the measurements over distances ranging from 10~m to 50~m. We also explored the use of passive reflectors to improve NLOS signal coverage at 28 GHz for both indoor and outdoor scenarios~\cite{wahab_indoor,wahab_outdoor}. The square metallic sheet reflectors were proved to be a simple, effective, and affordable way of enhancing mmWave coverage. With the proper shape and dimension of the reflector, the received power could be improved significantly, and it even approaches the Friis free space LOS received power at the same travel distance as the NLOS signal. 

From these previous studies, mmWave features a high path loss and a high material attenuation. A survey \cite{mahfuza2017} stated the demand of developing methodologies that support highly directional mmWave links over longer distances at airports due to the high path loss. Although the terrestrial application of mmWave systems is advancing at a rapid pace, the use of mmWave communication systems in aviation systems or airports is still in its infancy, partially due to the lack of characterization of mmWave wireless channels for the aviation field and the airport environment, and hence measurements in different airports are needed. A study \cite{mahfuza2018} conducted channel measurements at Boise Airport in both LOS and NLOS scenarios and presented a large scale fading channel model of 60 GHz. Another study \cite{mahfuza_vtc} focused on LOS mmWave propagation measurement and modeling also took place in Boise Airport.
To the best of authors' knowledge, only one study \cite{airport3bands} exists on airport mmWave propagation characterization at 28 GHz in an indoor scenario. This work focused on the specular propagation paths, specular and diffuse power contributions, polarization, and the delay and angular spreads at Helsinki Airport, Finland. Further results and analysis are needed at different airports (covering commercial aviation, private aviation, etc.) to develop better understanding of mmWave propagation in airport environments. 

\begin{figure*}[!t]
\centering
\includegraphics[width = .7\textwidth]{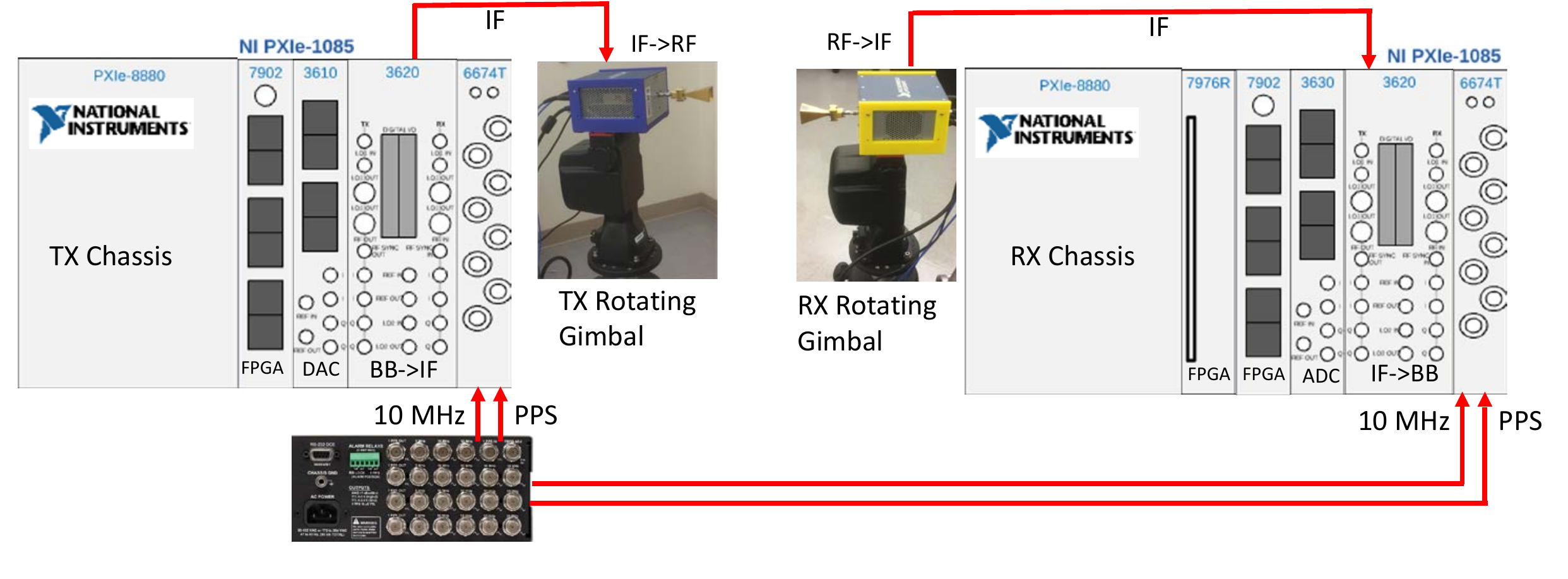}\vspace{-2mm}
\caption{Channel sounder block diagram for the NI PXI platform used in the experiments~\cite{erden2020correction}.} 
\label{Hardware_setup}\vspace{-4mm}
\end{figure*}

In this work, we performed mmWave channel sounding measurement for both indoor, outdoor and indoor-to-outdoor scenarios at 28 GHz in Johnston Regional Airport close to Raleigh, NC, based on NI's PXI-based channel sounding platform~\cite{NImmwave} (see Fig.~\ref{Hardware_setup}), using 17~dBi horn antennas on both gimbal-assisted transmitter (Tx) and receiver (Rx) to analyze propagation characteristics. A higher outdoor FSPL was observed comparing to indoor FSPL. Indoor propagation also showed more multipath components~(MPCs) and better signal coverage than the outdoor propagation in the airport environment, while indoor-to-outdoor propagation was difficult at 28 GHz.

\section{Propagation Modeling}

\subsection{LOS Propagation}

Path loss (PL) is the reduction in power density (attenuation) of an electromagnetic wave as it propagates through space. It is one of the key factors that impact the received signal strength and given by 
$\mathrm{PL}~\mathrm{(dB)} = P_{\rm t} - P_{\rm r}$,        
where $P_{\rm t}$ is the transmit power, and $P_{\rm r}$ is the received power. 
We consider the FSPL model, which yields: 
\begin{equation}
\mathrm{FSPL}~\mathrm{(dB)} =  20\log(d)+ 20\log(f)+20\log\left(\frac{4\pi}{c}\right)-G_{\rm T}-G_{\rm R}~,  \label{GrindEQ__4_}
\end{equation}
where $G_{\rm T}$ and $G_{\rm R}$ are Tx and Rx gain in the LOS direction, $d$ is the distance between Tx and Rx, and $f$ is the center frequency of the signal.

\subsection{ Reflection}

For reflection measurement, we consider the reflection coefficient $\Gamma$, which is the ratio of the amplitude of reflected signal to the amplitude of incident signal. Measured reflection coefficient is derived 
as follows \cite{ref_theory}:
\begin{equation}
|\Gamma| = \frac{d_{\rm total}}{d_{\rm LOS}} \sqrt\frac{P_{r_{\rm reflected}}}{P_{r_{\rm LOS}}}~,  \label{GrindEQ__5_} 
\end{equation}
\noindent where $d_{\rm LOS}$ is the distance between Tx and Rx for LOS propagation, and $d_{\rm total}$ is the total sum of travel distance of the incident and reflected ray. $P_{r_{\rm reflected}}$ is the received power of the reflected ray, and $P_{r_{\rm LOS}}$ is the LOS received power.

Theoretical reflection coefficient considers two conditions, which are perpendicular reflection coefficient $\Gamma_{\rm perpendicular}$ when E-field is perpendicular to the plane of incidence, and parallel reflection coefficient $\Gamma_{\rm parallel}$ when E-field is parallel to the plane of incidence:
\begin{align}
|\Gamma_{\rm perpendicular}| &= {\left|\frac{Z_2\cos\theta{_i}-Z_1\cos\theta{_t}}    {Z_2\cos\theta{_i}+Z_1\cos\theta{_t}}\right|}~,
\label{GrindEQ__6_}\\
|\Gamma_{\rm parallel}| &= {\left|\frac{Z_2\cos\theta{_t}-Z_1\cos\theta{_i}}    {Z_2\cos\theta{_t}+\eta{_1}\cos\theta{_i}}\right|}~, 
\label{GrindEQ__7_}
\end{align}
where $Z_1$ and $Z_2$ are the wave impedance of media 1 and 2, $\theta{_i}$ is the incident angle, and $\theta{_t}$ is the refracted angle. Assuming that the media are non-magnetic (i.e., relative permeability $\mu{_r}$ = 1), the wave impedance is determined solely by the refractive index $\eta$. We can further substitute wave impedance to refractive index using \eqref{GrindEQ__Z_} and get \eqref{GrindEQ__66_} and \eqref{GrindEQ__77_}:
\begin{align}
Z_i&=\frac{Z_0}{\eta{_i}}~,
\label{GrindEQ__Z_}\\
|\Gamma_{\rm perpendicular}| &= {\left|\frac{\eta{_2}\cos\theta{_i}-\eta{_1}\cos\theta{_t}}    {\eta{_2}\cos\theta{_i}+\eta{_1}\cos\theta{_t}}\right|}~,
\label{GrindEQ__66_}\\
|\Gamma_{\rm parallel}| &= {\left|\frac{\eta{_2}\cos\theta{_t}-\eta{_1}\cos\theta{_i}}    {\eta{_2}\cos\theta{_t}+\eta{_1}\cos\theta{_i}}\right|}~,
\label{GrindEQ__77_}
\end{align}
where $\eta{_1}$ and $\eta{_2}$ are the refractive indices for media 1 and 2. Using Snell's law we can get the relationship of angle versus refractive index: 
\begin{equation} 
\label{GrindEQ__8_} 
\frac{\sin{\theta{_i}}}{\sin{\theta{_t}}} = \frac{\eta{_2}}{\eta{_1}}~. 
\end{equation} 
We can also obtain the relationship between refractive index and frequency dependent relative permittivity $\epsilon{_r}(f)$ using:
\begin{equation} 
\label{GrindEQ__9_} 
\eta=\sqrt{\epsilon{_r}(f)\mu{_r}}=\sqrt{\epsilon{_r}(f)}~.
\end{equation} 
Since $\eta{_1}$ is the refraction index of air, which is equal to 1, we can substitute \eqref{GrindEQ__8_} and \eqref{GrindEQ__9_} into \eqref{GrindEQ__66_} and \eqref{GrindEQ__77_} to further simplify it as follows~\cite{derive}: 
\begin{align}
|\Gamma_{\rm perpendicular}| &= {\left|\frac{\cos\theta{_i} - \sqrt{\epsilon{_r}(f)-\sin^2\theta{_i}}}    {\cos\theta{_i} + \sqrt{\epsilon{_r}(f)-\sin^2\theta{_i}}}\right|}~, 
\label{GrindEQ__10_}\\
|\Gamma_{\rm parallel}| &= {\left|\frac{\epsilon{_r}(f)\cos\theta{_i} - \sqrt{\epsilon{_r}(f)-\sin^2\theta{_i}}}  {\epsilon{_r}(f)\cos\theta{_i} + \sqrt{\epsilon{_r}(f)-\sin^2\theta{_i}}}\right|}~. 
\label{GrindEQ__11_}
\end{align}
Note that the equations above contain only two parameters: incident angle and relative permittivity (frequency and material dependent). Due to our measurement setup, we only consider parallel condition for our analysis and take the value of $\epsilon{_r}=3$ based on our previous measurement of clear glass relative permittivity at 28 GHz in an indoor environment at NC State University.

\subsection{Penetration}

For penetration loss, we do not have specific model as our theoretical formula of penetration loss. Therefore, we will transmit mmWave signal through wall and glass door and compare its loss with the LOS condition. We can calculate the measured penetration loss by: 
\begin{equation}
\mathrm{Penetration~Loss~(dB)} = P_{r_{\rm LOS}} - P_{r_{\rm penetration}}~.  \label{GrindEQ__12_}
\end{equation}





\begin{figure}[!t]
\centering
\includegraphics[width = .425\textwidth]{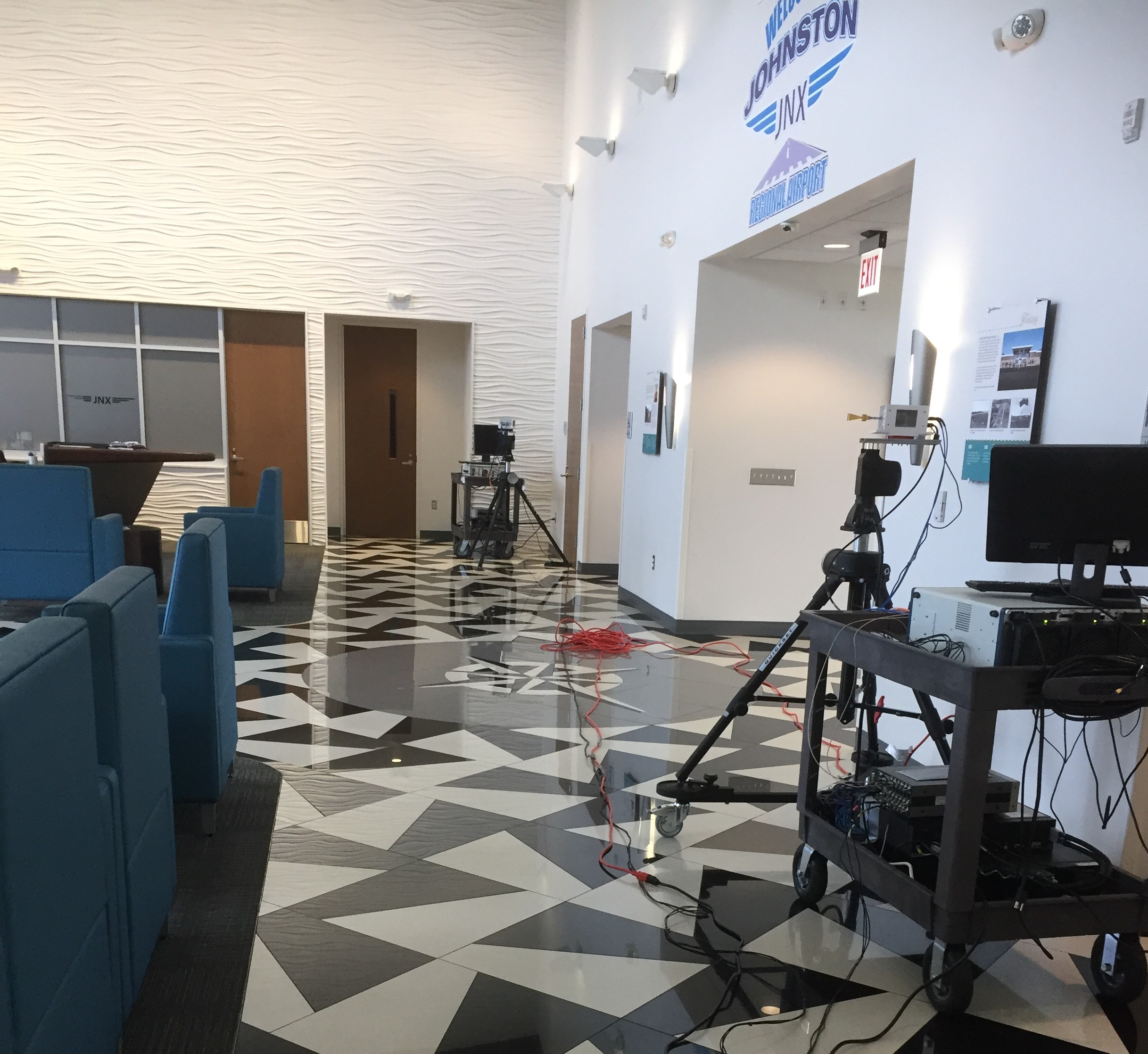}
\caption{Indoor measurement at Johnston Regional Airport.} 
\label{Indoor}\vspace{-4mm}
\end{figure}

\section{Measurement Setup}

Our channel sounder hardware is based on NI's mmWave system at 28~GHz, as shown in Fig.~\ref{Hardware_setup}~\cite{erden2020correction}. It consisted of NI PXIe-1085 Tx and Rx chassis, 28 GHz Tx and Rx mmWave radio heads (fixed on FLIR~PTU-D48E gimbals) from NI, and FS725 Rubidium (Rb) clocks. A 10~MHz pulse per second (PPS) signals generated by a single clock~\cite{SRS} was connected to PXIe 6674T modules at both Tx and Rx. The 10~MHz signal was used to generate the required local oscillator~(LO) signals, and the transmission at TX side and reception at the RX side were triggered by the same PPS signal.

The sounder code was based on LabVIEW, and periodically transmitted a Zadoff-Chu (ZC) sequence of length 2048 to sound the channel. It was then filtered by the root-raised-cosine (RRC) filter, and the generated samples were uploaded to PXIe-7902 FPGA. These samples were sent to PXIe-3610 digital-to-analog (DAC) converter with a sampling rate of $f_{s}=3.072$~GS/s. The PXIe-3620 module up-converted the base-band signal to IF, and the 28~GHz mmWave radio head up-converted the IF signal to RF with 2 GHz bandwidth and 10~dBm transmit power. At the Rx, 28 GHz mmWave radio head down-converted the RF signal to IF, which was then down-converted to base-band at the PXIe-3620. The PXIe-3630 analog-to-digital converter (ADC) sampled the base-band analog signal with the sampling rate of $f_{s}=3.072$~GS/s. 

The channel sounder provides $2/f_{s}=0.651$~ns delay resolution in the time domain. Therefore, any multipath component with a delay higher than $0.651$~ns can be resolved, which represent a path-length difference of $0.195$~m. The correlation and averaging operations were performed in PXIe-7976R FPGA operation, and the complex CIR samples and the power-delay-profile (PDP) were sent to the PXIe-8880 host PC for further processing and saving to local disk. Calibration and equalization was then performed to eliminate the channel distortion due to the non-idealities of the hardware themselves. After that, directional horn antennas were connected to the mmWave radio heads at the Tx and the Rx sides with 17~dBi gains, and 26 degree and 24 degree half power beam-widths in the elevation and azimuth planes, respectively. For all the measurements, both Tx and Rx scanned an azimuth degree of $-167.98$ to $-167.98$ with a resolution of $20$ degree at each Tx-Rx position.

\begin{figure}[!t]
\centering
\includegraphics[width = .45\textwidth]{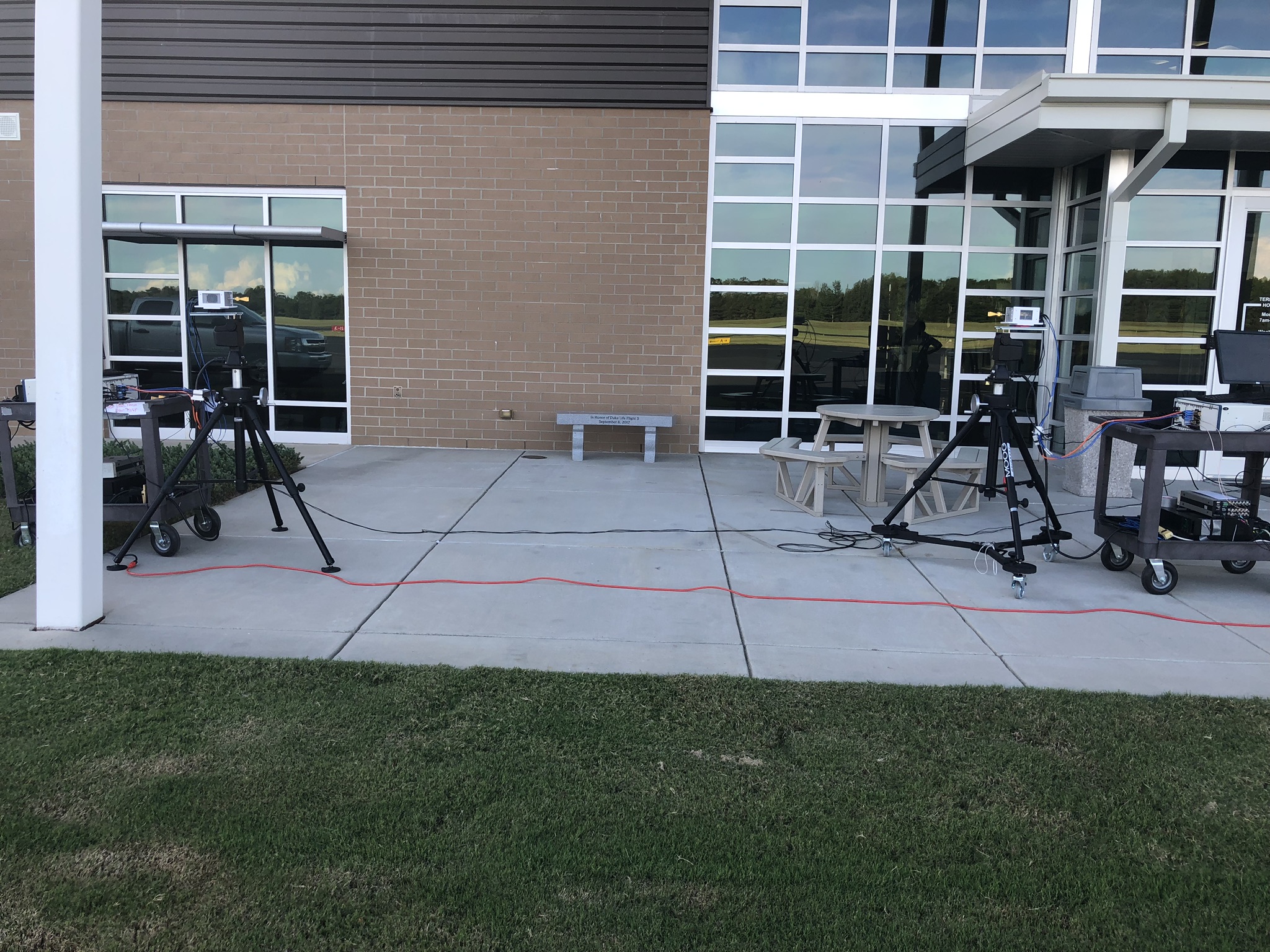}
\caption{Outdoor measurement at Johnston Regional Airport.} 
\label{Outdoor}\vspace{-4mm}
\end{figure}

\subsection{Indoor Measurement Setup}

As shown in Fig.~\ref{Indoor}, for the indoor measurements, Tx and Rx were placed inside the terminal hall at the same height of 1.5~m and were aligned to each other. Tx was fixed at one position, and the received PDP of Rx was measured at separation of 5, 10, and 15~m from Tx. 

\subsection{Outdoor Measurement Setup}

\begin{figure*}[!t]
\centering
\includegraphics[trim=0cm 1.5cm 0cm 1.7cm, clip,width = .73\textwidth]{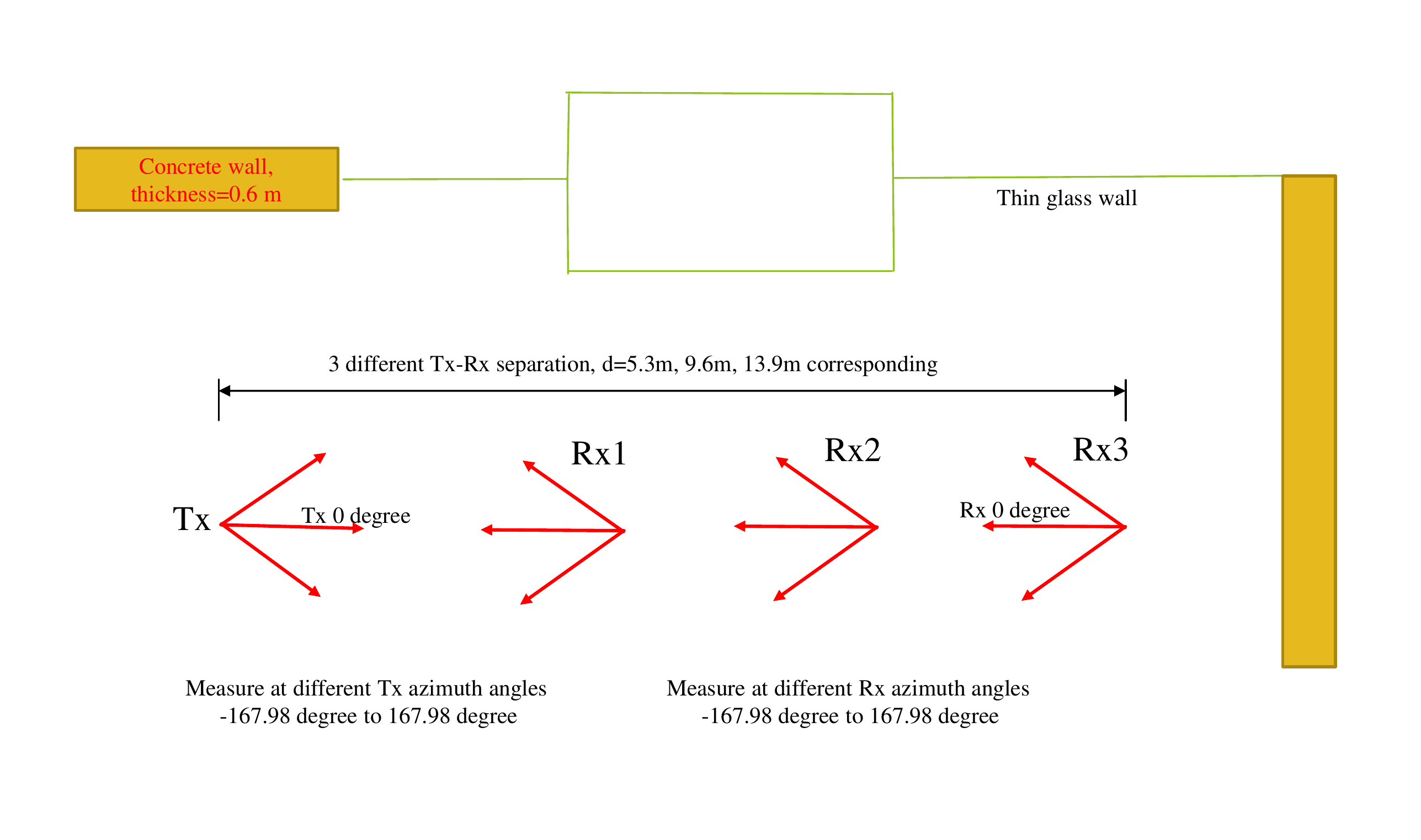}
\caption{Outdoor measurement scenario (looking from above).} \vspace{-4mm}
\label{Out_out}
\end{figure*}

In the outdoor case shown in Fig.~\ref{Outdoor}, the Tx and the Rx were placed outside the airport terminal. The Tx was fixed, and the Rx was aligned to the Tx with a separation of 5.3, 9.6, and 13.9~m from the Tx, as shown in Fig.~\ref{Out_out}.

\subsection{Indoor-to-Outdoor Measurement Setup}

\begin{figure*}[!t]
\centering
\includegraphics[trim=0cm 1.5cm 0cm 1.3cm, clip,width = .73\textwidth]{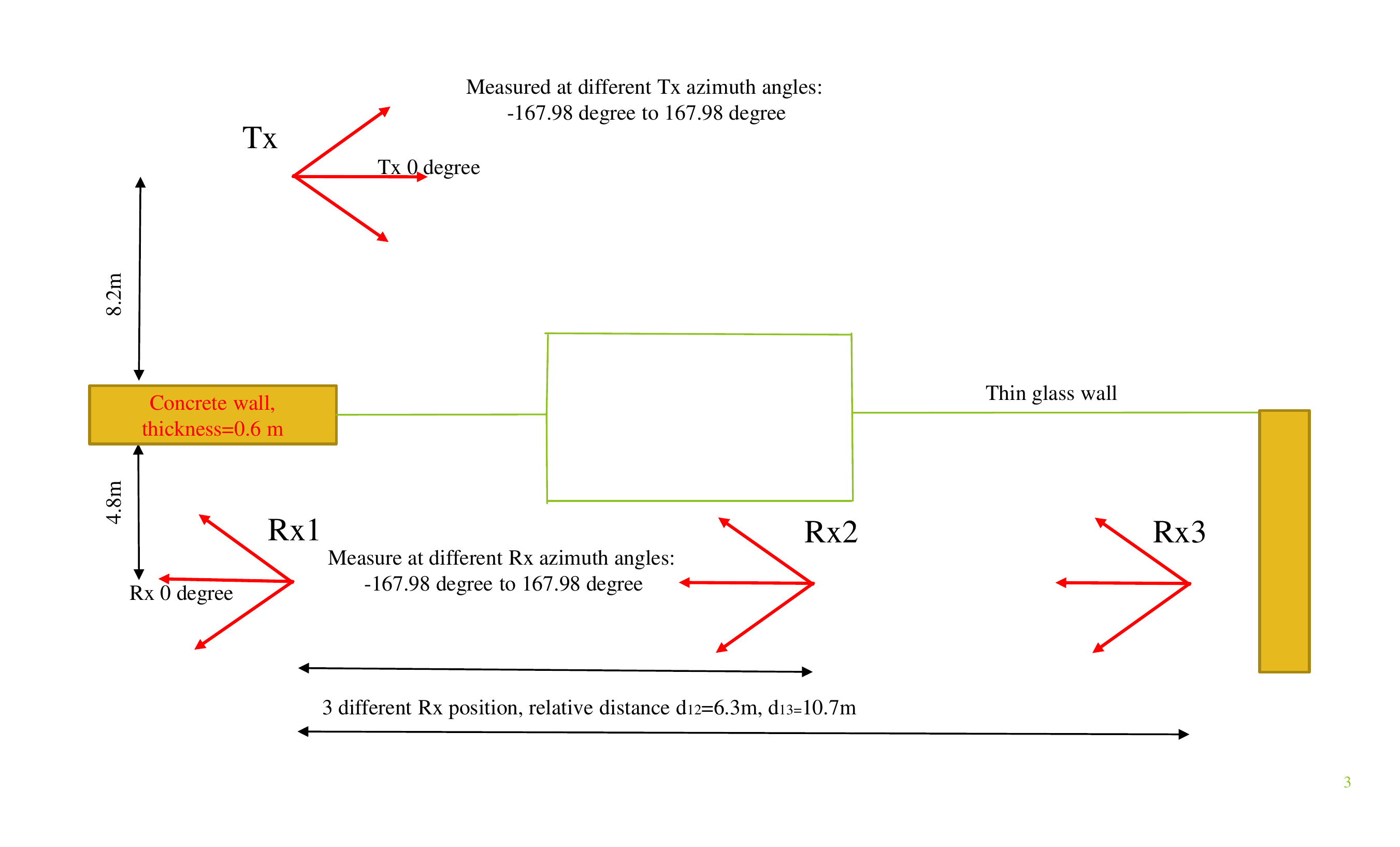}
\caption{Indoor-to-outdoor measurement scenario (looking from above).} \vspace{-4mm}
\label{In_out}
\end{figure*}

In the indoor-to-outdoor measurement scenario, Tx was fixed inside the terminal hall, and Rx was placed outside at a separation of 13.4~m. The Rx would later move along the wall at distance of 6.3~m and 10.7~m away from its first position, which is shown in Fig.~\ref{In_out}. The direct LOS propagation was blocked by the terminal wall and clear glass door, of which between Tx and Rx1 was a concrete wall, between Tx and Rx2 and Tx and Rx3 was a glass wall.

\section{Results and analysis}

\begin{figure}[!t]
\centering
\includegraphics[width = .48\textwidth]{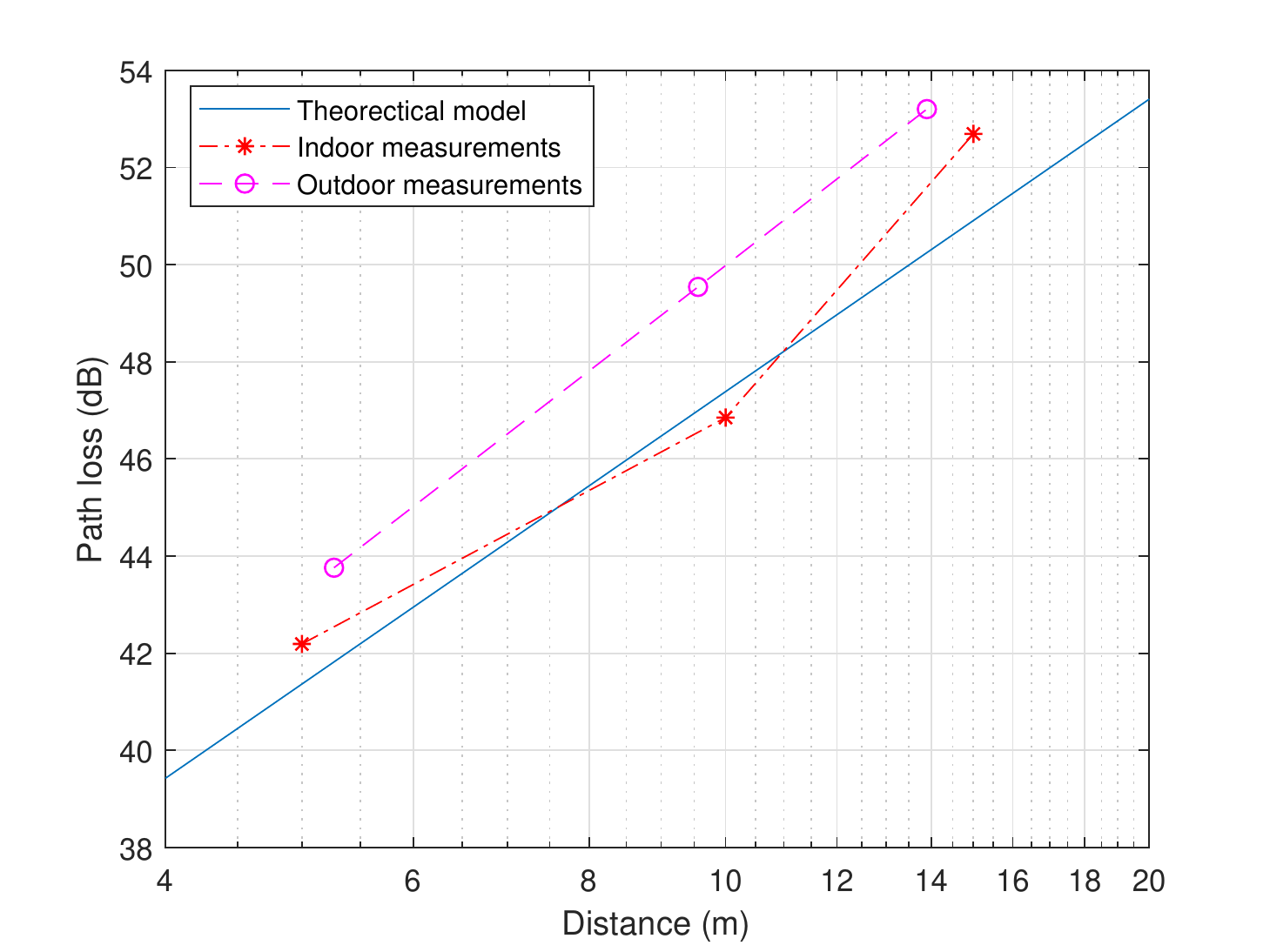}
\caption{FSPL at 28~GHz. Blue line is the theoretical FSPL, and red asterisk is the measured indoor FSPL at 5, 10, and 15~m. Purple circle is the measured outdoor FSPL at 5.3, 9.6, and 13.9~m.} 
\label{result_LOS}
\end{figure}

\subsection{LOS Propagation}

We measured the LOS path loss in both indoor and outdoor scenarios. Each measured point was taken when Tx and Rx were aligned to each other at a certain Tx-Rx separation. As shown in Fig.~\ref{result_LOS}, the theoretical FSPL calculated from \eqref{GrindEQ__4_} is further compared with the measured path-loss.
From Fig.~\ref{result_LOS}, indoor path loss matches the theoretical FSPL. However outdoor path loss is approximately 2-3~dB higher than indoor path loss. This might be a consequence of a more complicated and unstable outdoor environment (wind, temperature, humidity, environment noise, etc.) compared to indoor.

\begin{figure}[!t]
\centering\vspace{-2mm}
\includegraphics[trim=0cm 0cm 0cm 0.8cm, clip,width = .48\textwidth]{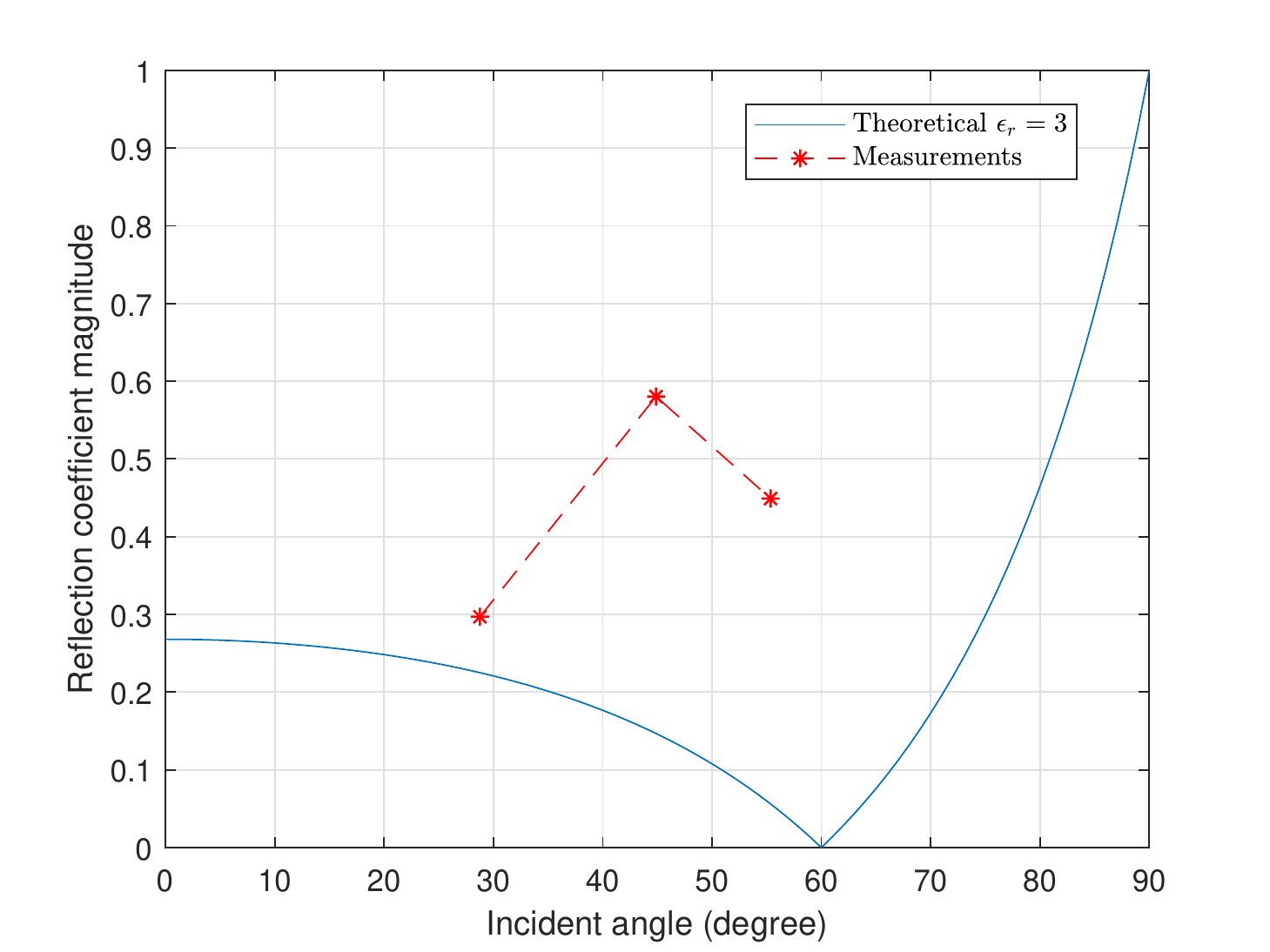}\vspace{-2mm}
\caption{Reflection coefficient of theoretical derivation and measurement result. Blue line indicates the theoretical value, and the dashed line with red asterisk is the measured coefficient at incident angle of $28.77$, $44.88$, and $55.36$ degree. For theoretical plot, the antenna radiation pattern is neglected.} 
\label{result_reflection}\vspace{-2mm}
\end{figure}

\subsection{Reflection}

The first-order reflection from the terminal glass wall and glass door at 3 different distances (which represent 3 different incident angles) from the outdoor measurement results were taken for further analysis. They are also compared with the theoretical reflection coefficient, as shown in Fig.~\ref{result_reflection}, of parallel condition (vertical polarized wave for our horn antennas) calculated via \eqref{GrindEQ__11_} at a relative permittivity of 3 for clear glass at 28~GHz. 

As seen in Fig.~\ref{result_reflection}, the measured results do not align with the theoretical calculation under the given assumptions. We have neglected the specific radiation pattern of the Tx/Rx antennas and the associated antenna gains for the Tx/Rx signal that correspond to the reflection point, and further work is needed to accurately model this artifact. 
The type of glass for the wall and door are not available to us and the ranging of relative permitivity for different types of glass could be ranging from 2-10 \cite{dialectric}. More thorough study needs to be conducted for different types of glass at different mmWave bands. Notably, the measurement environment (the double-layer glass door) might also introduce some unresolvable MPCs due to scattering and higher-order reflections which may lead to a lower observed reflection loss.

\begin{figure*}[t]
 \centering
	\begin{subfigure}{0.66\columnwidth}
	\centering
	\includegraphics[width=\columnwidth]{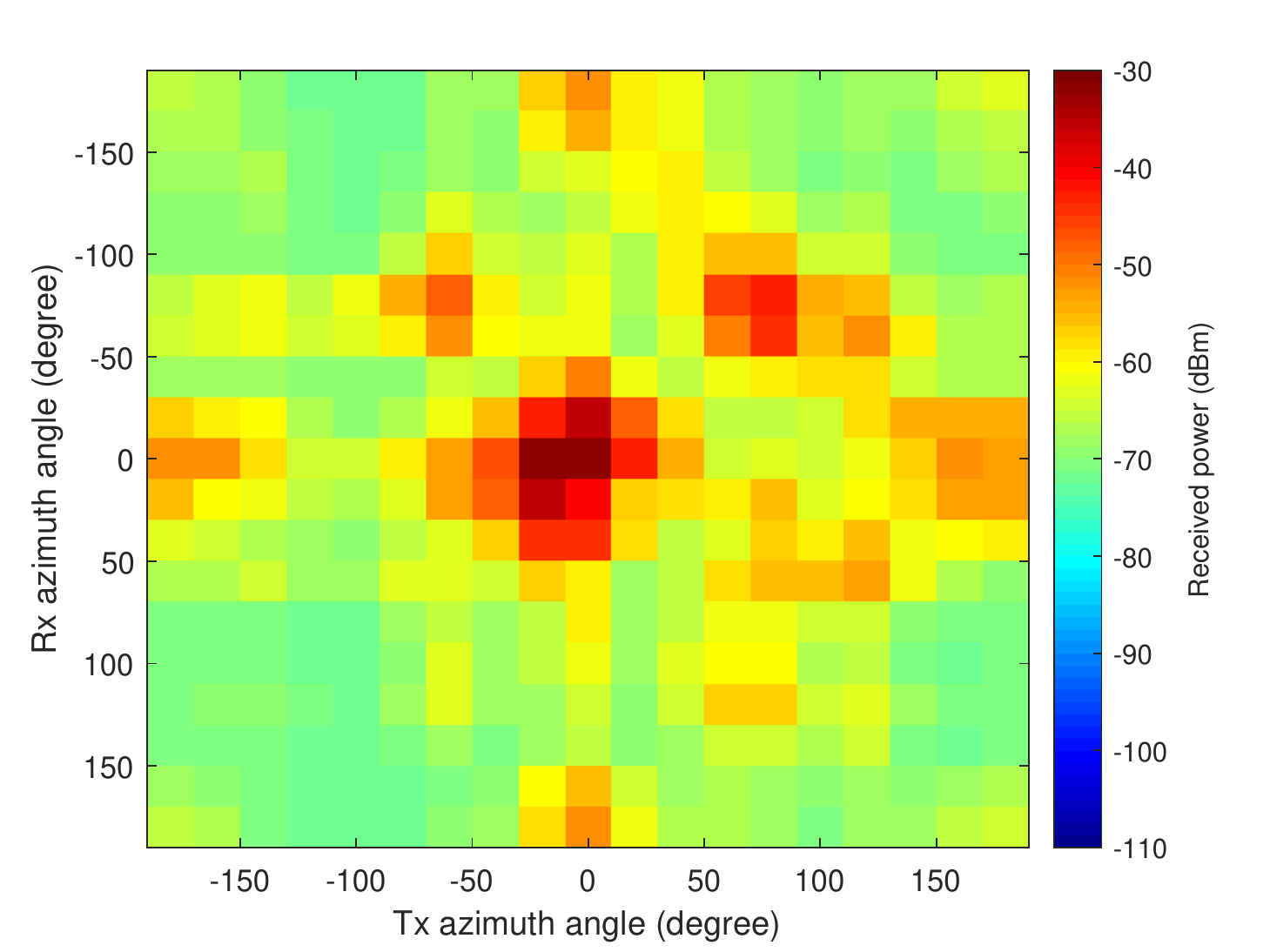} 
	\caption{}
    \end{subfigure}			
	\begin{subfigure}{0.66\columnwidth}
	\centering
    \includegraphics[width=\columnwidth]{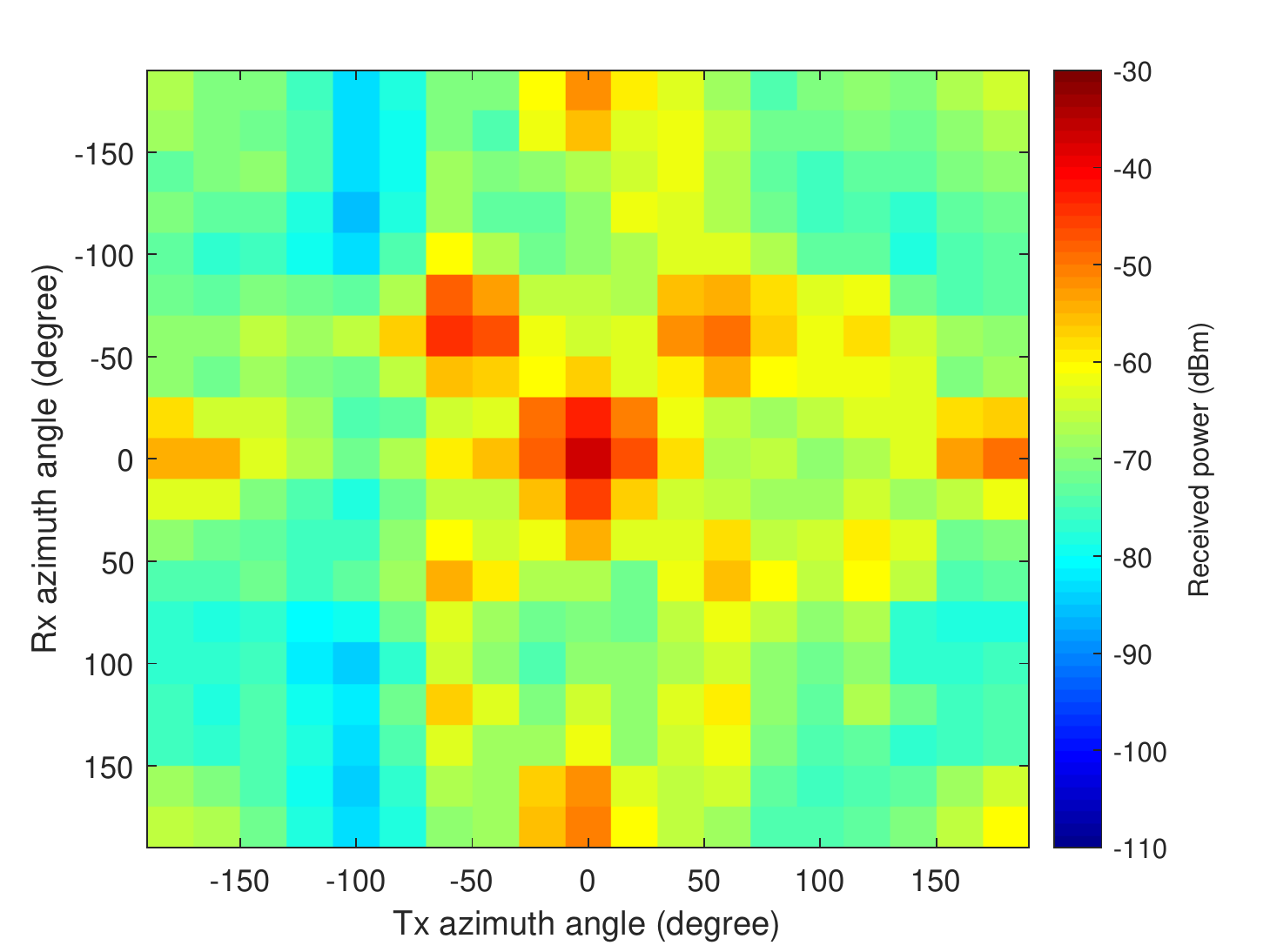}
	\caption{}
    \end{subfigure}
    \begin{subfigure}{0.66\columnwidth}
	\centering
	\includegraphics[width=\columnwidth]{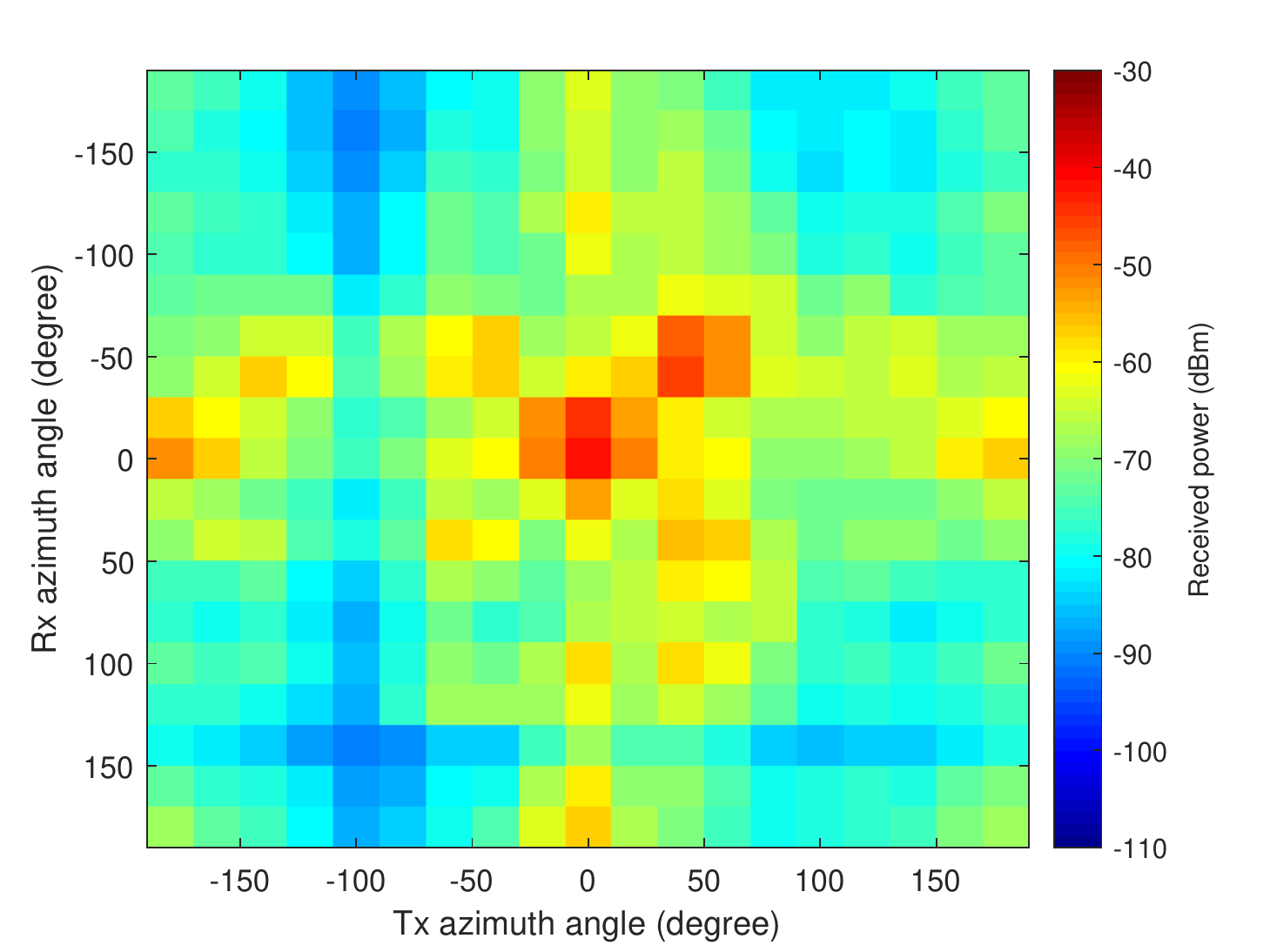} 
	\caption{}
    \end{subfigure}			
	\begin{subfigure}{0.66\columnwidth}
	\centering
    \includegraphics[width=\columnwidth]{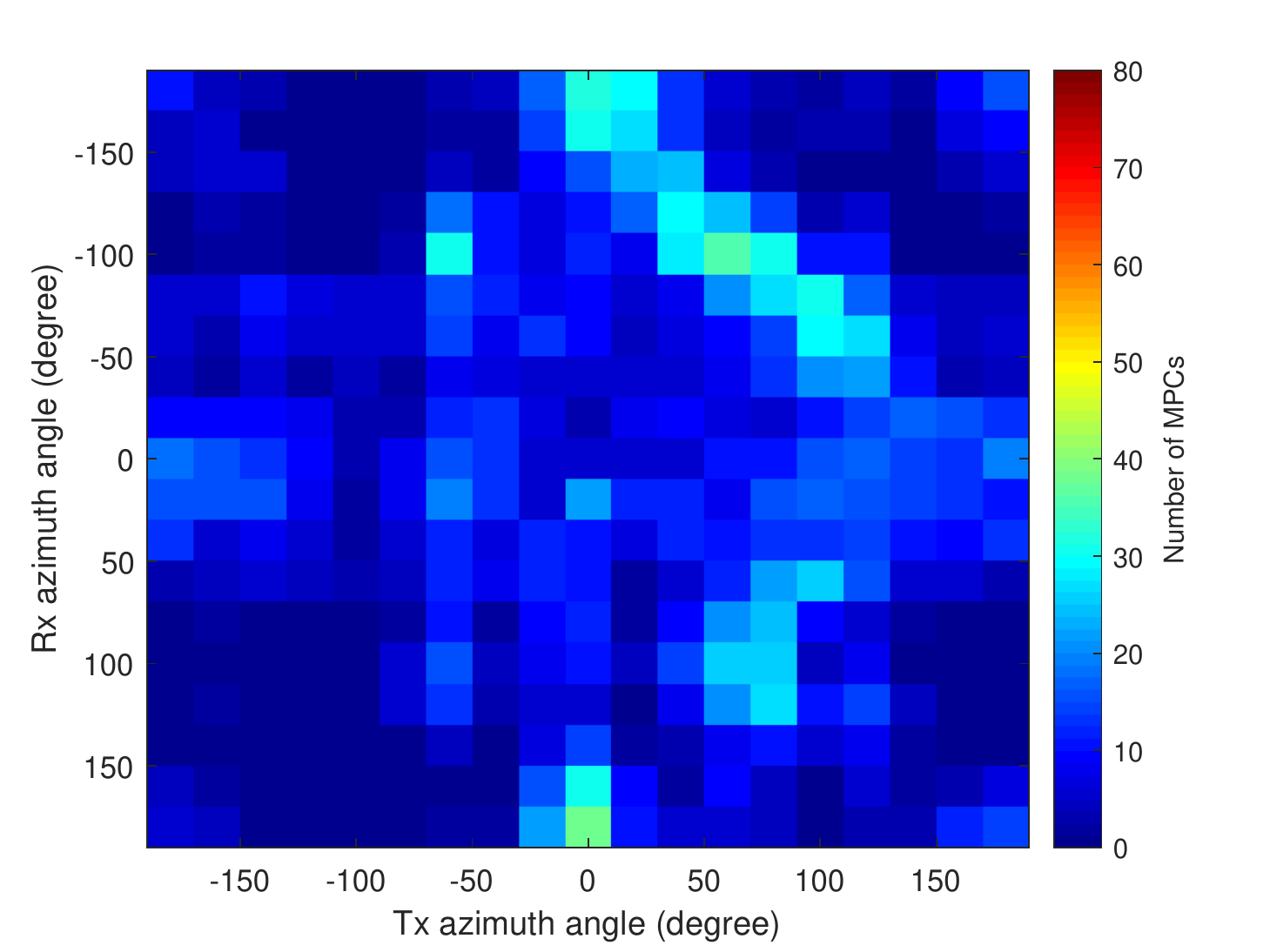}
	\caption{}
    \end{subfigure}
    \begin{subfigure}{0.66\columnwidth}
	\centering
	\includegraphics[width=\columnwidth]{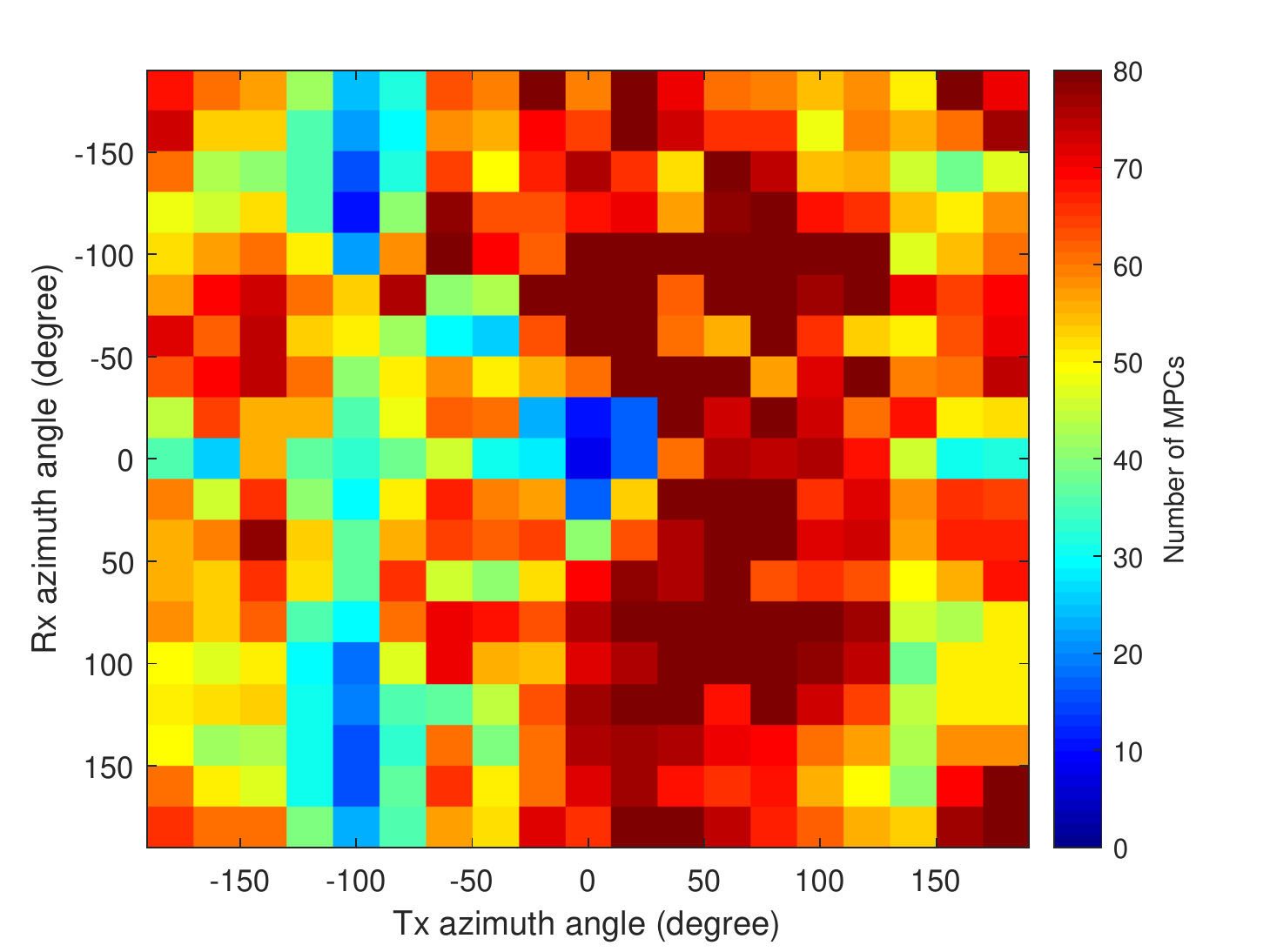} 
	\caption{}
    \end{subfigure}			
	\begin{subfigure}{0.66\columnwidth}
	\centering
    \includegraphics[width=\columnwidth]{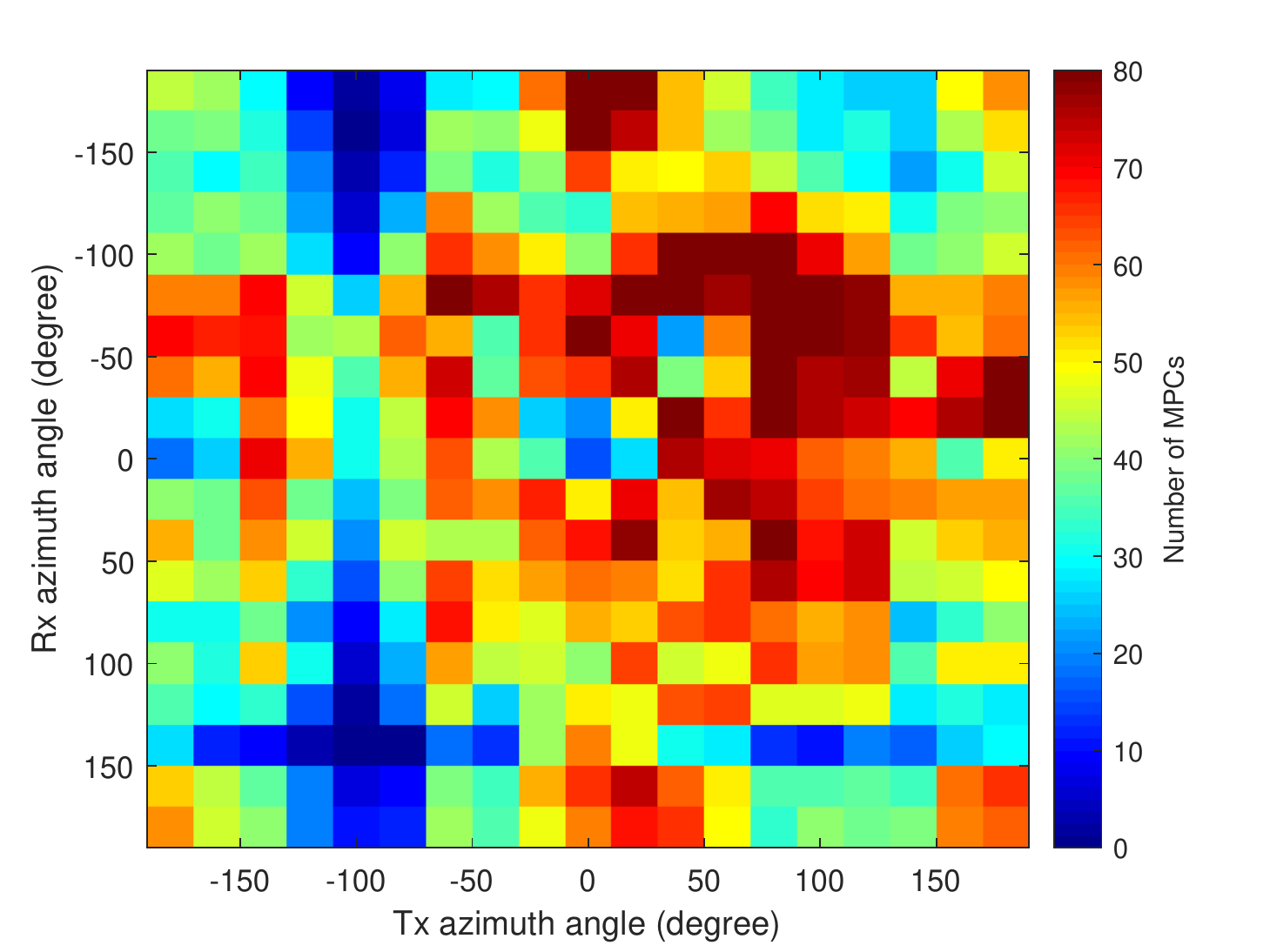}
	\caption{}
    \end{subfigure}
    \caption{Indoor received power at different Tx and Rx azimuth angle for a Tx-Rx separation of (a) 5~m, (b) 10~m, (c) 15~m. Number of MPCs at different Tx and Rx azimuth angle for a Tx-Rx separation of (d) 5~m, (e) 10~m, (f) 15~m. The angular resolution at the Tx and the Rx is $20$ degrees.}
    \label{in_in_result}
    \vspace{-3mm}
\end{figure*}

\subsection{Penetration}

A comparison of concrete wall and glass wall penetration loss with the reference measurement of concrete blocks and glass board \cite{hosseini2020attenuation} are shown in Table~\ref{Table}. Path loss without blockage is the calculated FSPL of the LOS MPC (at the same travel distance as the blocked case) from \eqref{GrindEQ__4_}. Path loss with blockage is the measured indoor-to-outdoor path loss of the blocked LOS ray. Penetration loss is calculated from \eqref{GrindEQ__12_} based on the difference of path loss with and without blockage. The attenuation factor in dB/cm is the averaged penetration loss over thickness.
The results show that the large thickness of concrete wall and the high attenuation factor of glass led to high penetration loss for both walls. The concrete attenuation factor matched the reference result. The slight larger attenuation is due to the outdoor propagation of the penetrated rays: since the reference measurement was taken indoor, a higher attenuation compared to the reference result was expected. For the attenuation result of glass in this measurement, the incident ray is not perpendicular to the glass wall and the effective thickness should be higher than the measured thickness, which makes it higher than the reference result.

\begin{table}
\footnotesize
\centering
\caption{Penetration Loss Measurement at 28 GHz.} 
\label{Table}
\begin{tabular}
{|p{1.8in}|p{0.4in}|p{0.4in}|} \hline \centering 
 \textbf{Parameter} & \textbf{Concrete} & \textbf{Glass} \\ \hline 
Path loss without blockage (dB) & 50.90 & 52.69 \\ \hline 
Path loss with blockage (dB) & 119.76 & 97.00 \\ \hline 
Penetration loss (dB) & 68.86 & 47.31 \\ \hline 
Thickness (cm) & 60 & 10 \\ \hline 
Measured attenuation factor (dB/cm) & 1.15 & 4.73 \\ \hline 
Reference attenuation factor (dB/cm)~\cite{hosseini2020attenuation} & 1.06 & 4.39 \\ \hline 
\end{tabular}
\end{table}

\subsection{Signal Coverage and MPCs}
The signal coverage and MPC distribution results for indoor measurement are shown in Fig.~\ref{in_in_result}. Each colored grid represents the received power or the number of MPCs at a given combination of Tx and Rx azimuth angle. As the Tx and the Rx separation got larger, the number of MPCs increases and its distribution varies a lot. This is because in the indoor scenario, as the distance increases, more reflective objects (e.g. chair, desk, etc.) in the TX and the RX surroundings are introduced to the channel, which results in an increase in the number of MPCs. Also, there is a dominant LOS propagation (for the strongest MPC) with azimuth angle of arrival (AoA) and angle of departure (AoD) both at 0 degree, together with several strong NLOS MPCs whose received power get close to the LOS MPC's received power.

\begin{figure*}[h]
\centering
	\begin{subfigure}{0.66\columnwidth}
	\centering
	\includegraphics[width=\columnwidth]{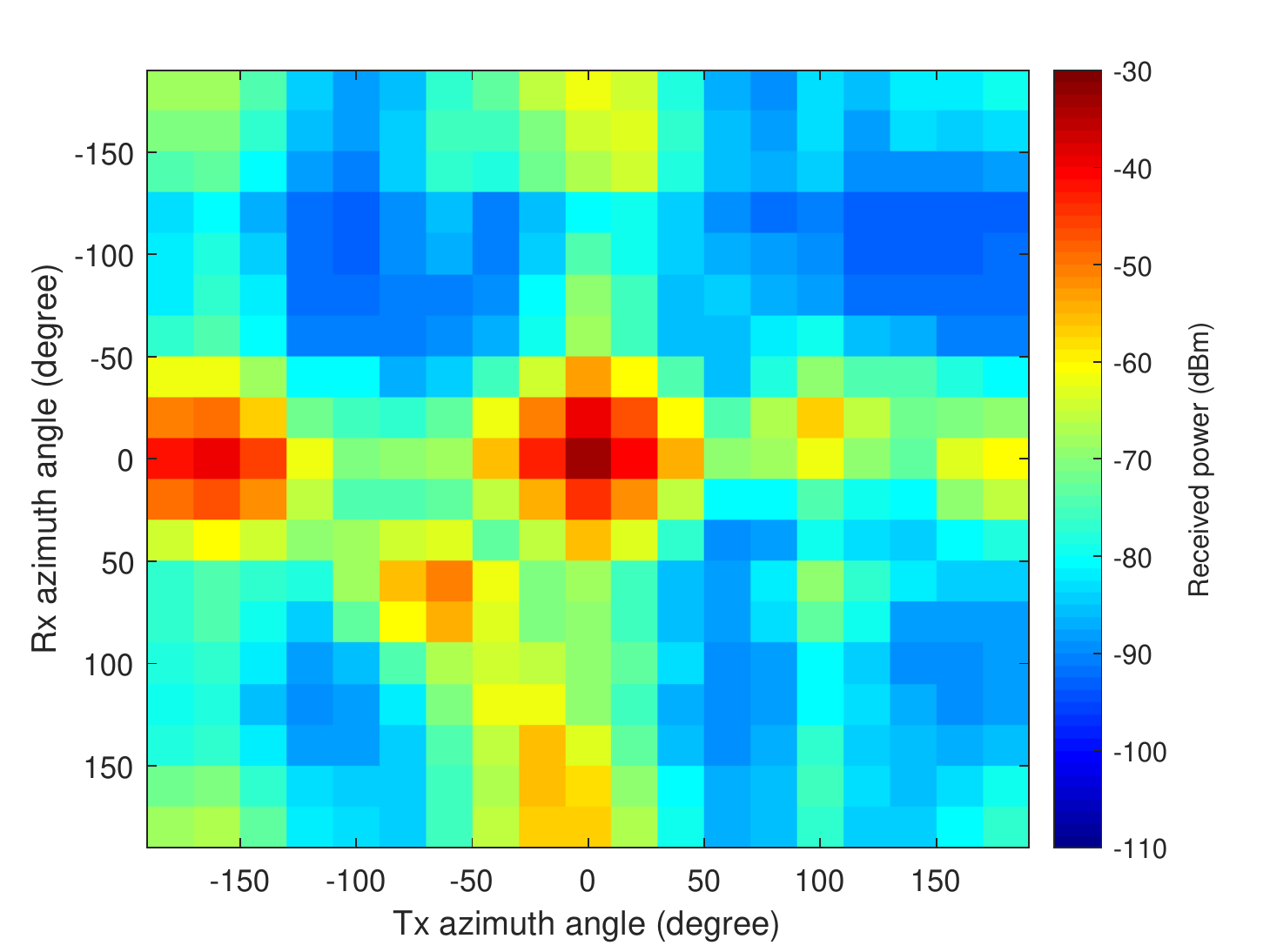} 
	\caption{}
    \end{subfigure}	
	\begin{subfigure}	{0.66\columnwidth}	
	\centering
    \includegraphics[width=\columnwidth]{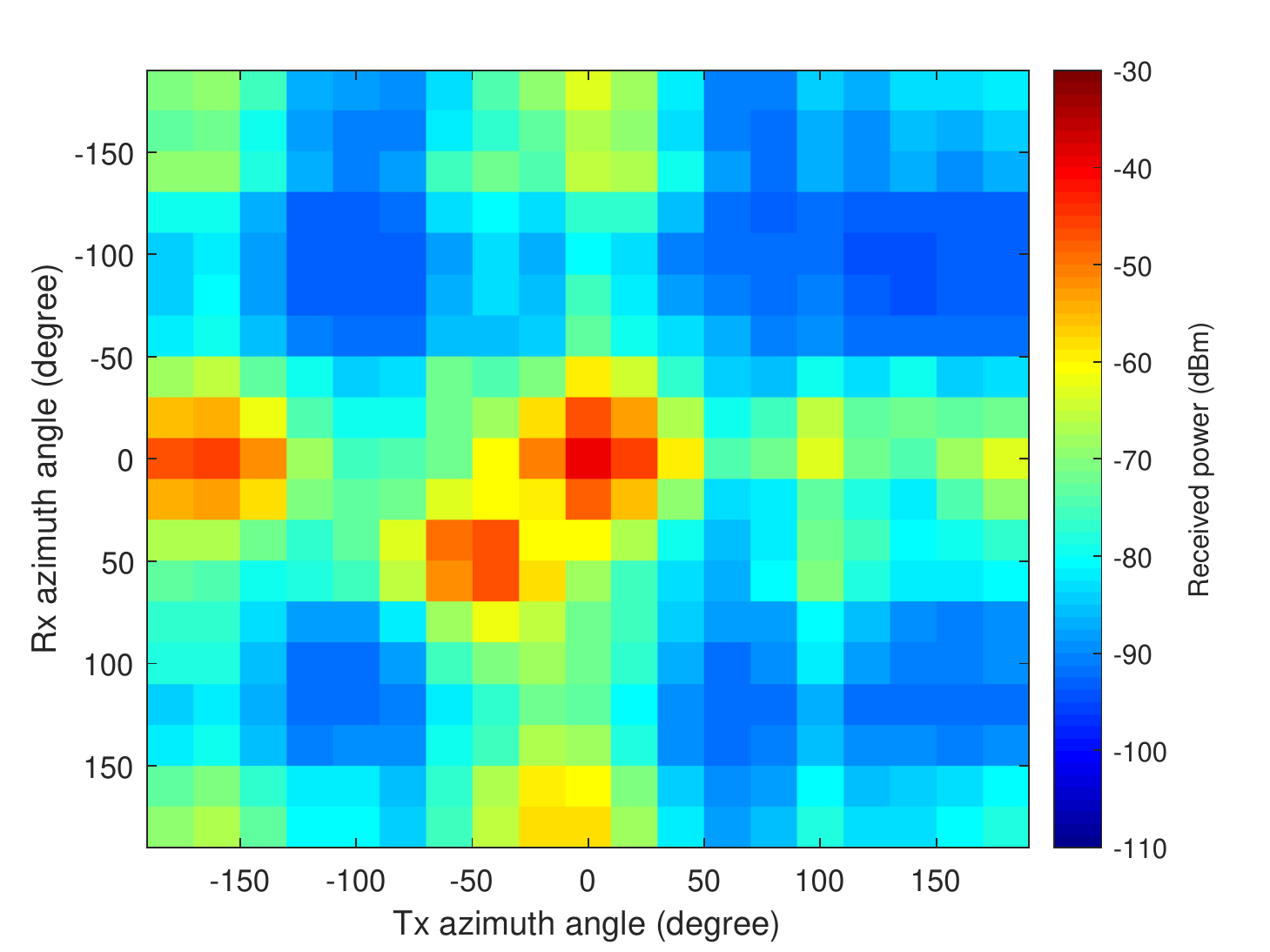}
	\caption{}
    \end{subfigure}
    \begin{subfigure}{0.66\columnwidth}
	\centering
	\includegraphics[width=\columnwidth]{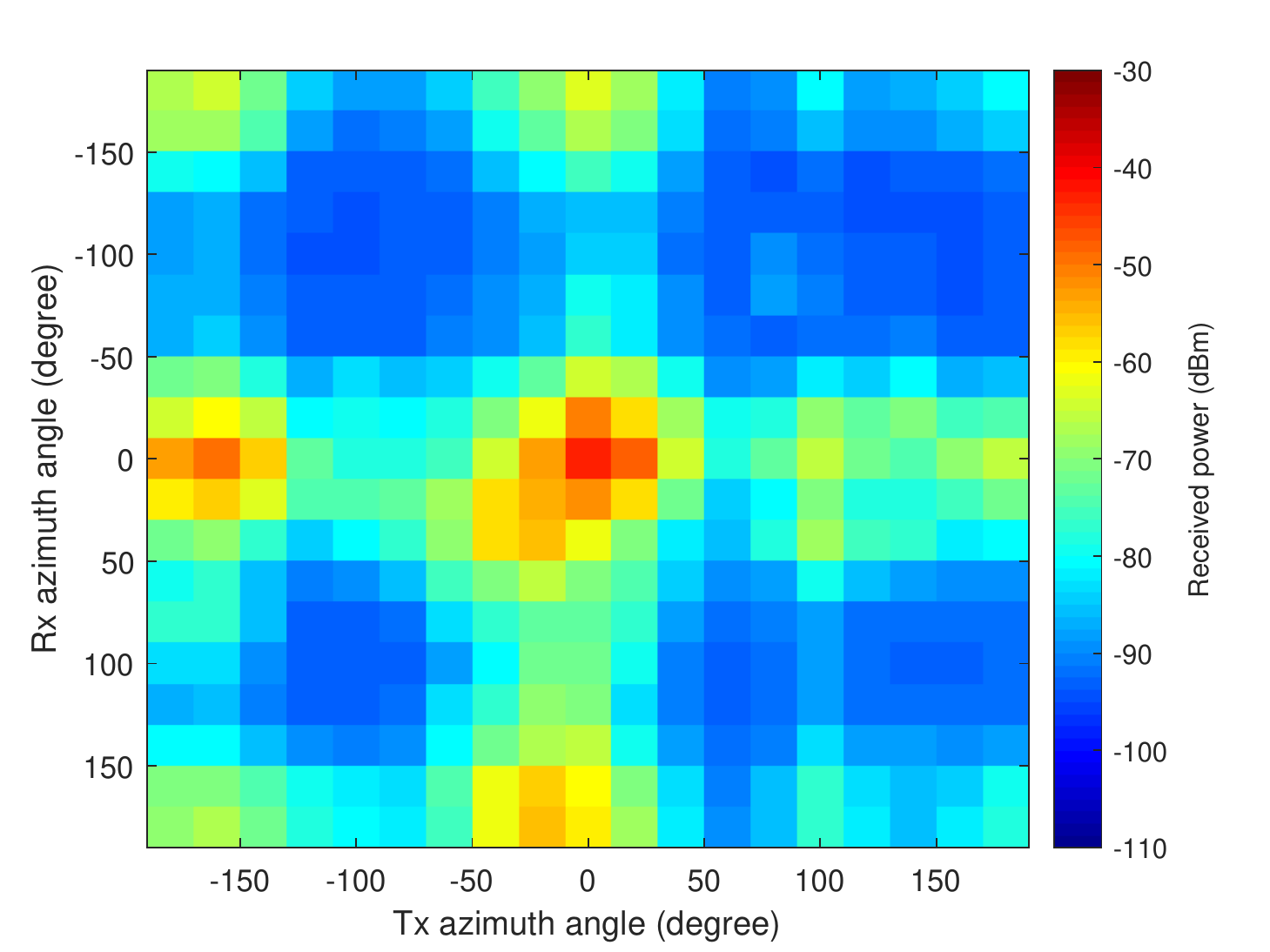} 
	\caption{}
    \end{subfigure}			
	\begin{subfigure}{0.66\columnwidth}
	\centering
    \includegraphics[width=\columnwidth]{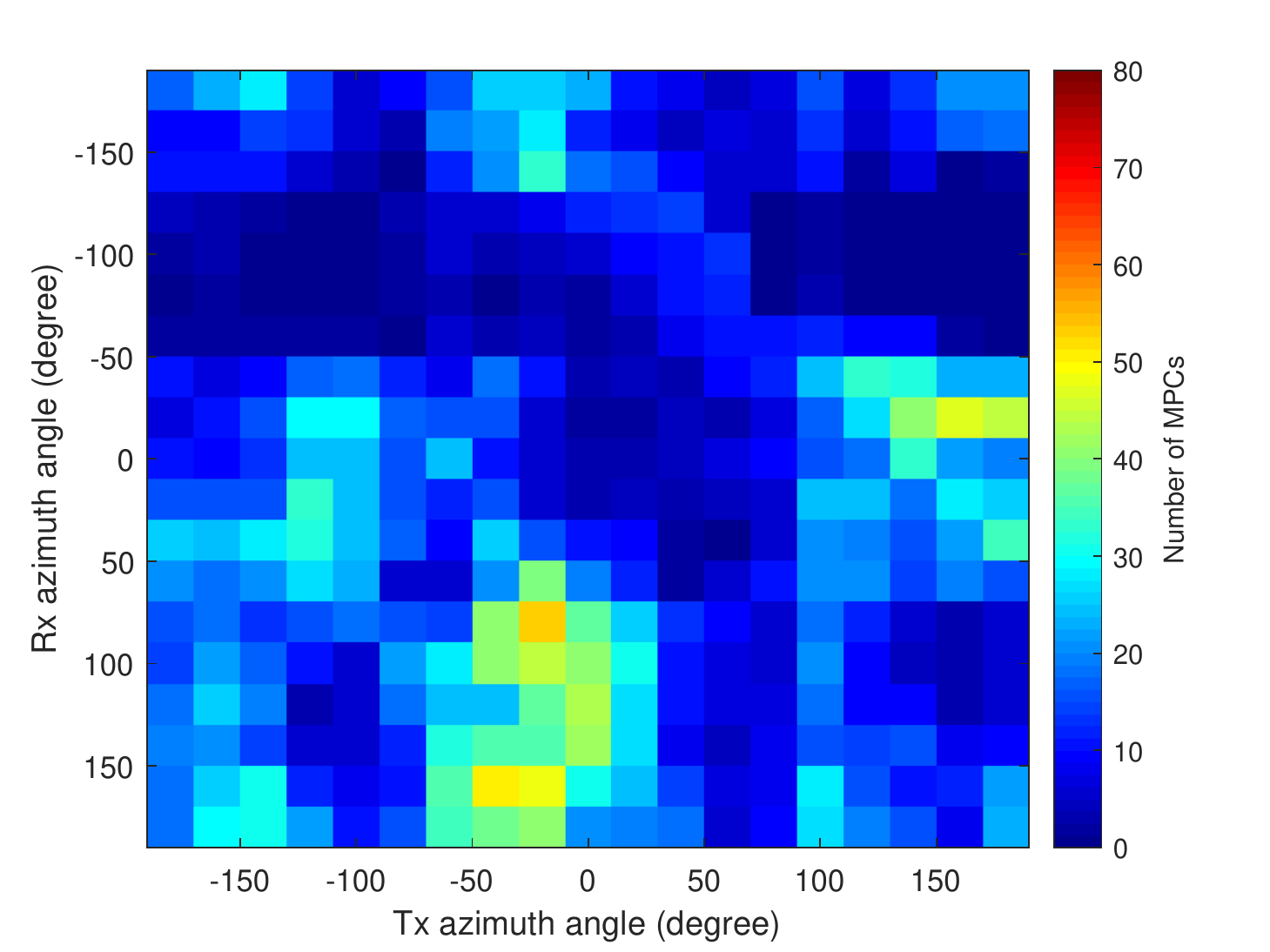}
	\caption{}
    \end{subfigure}
    \begin{subfigure}{0.66\columnwidth}
	\centering
	\includegraphics[width=\columnwidth]{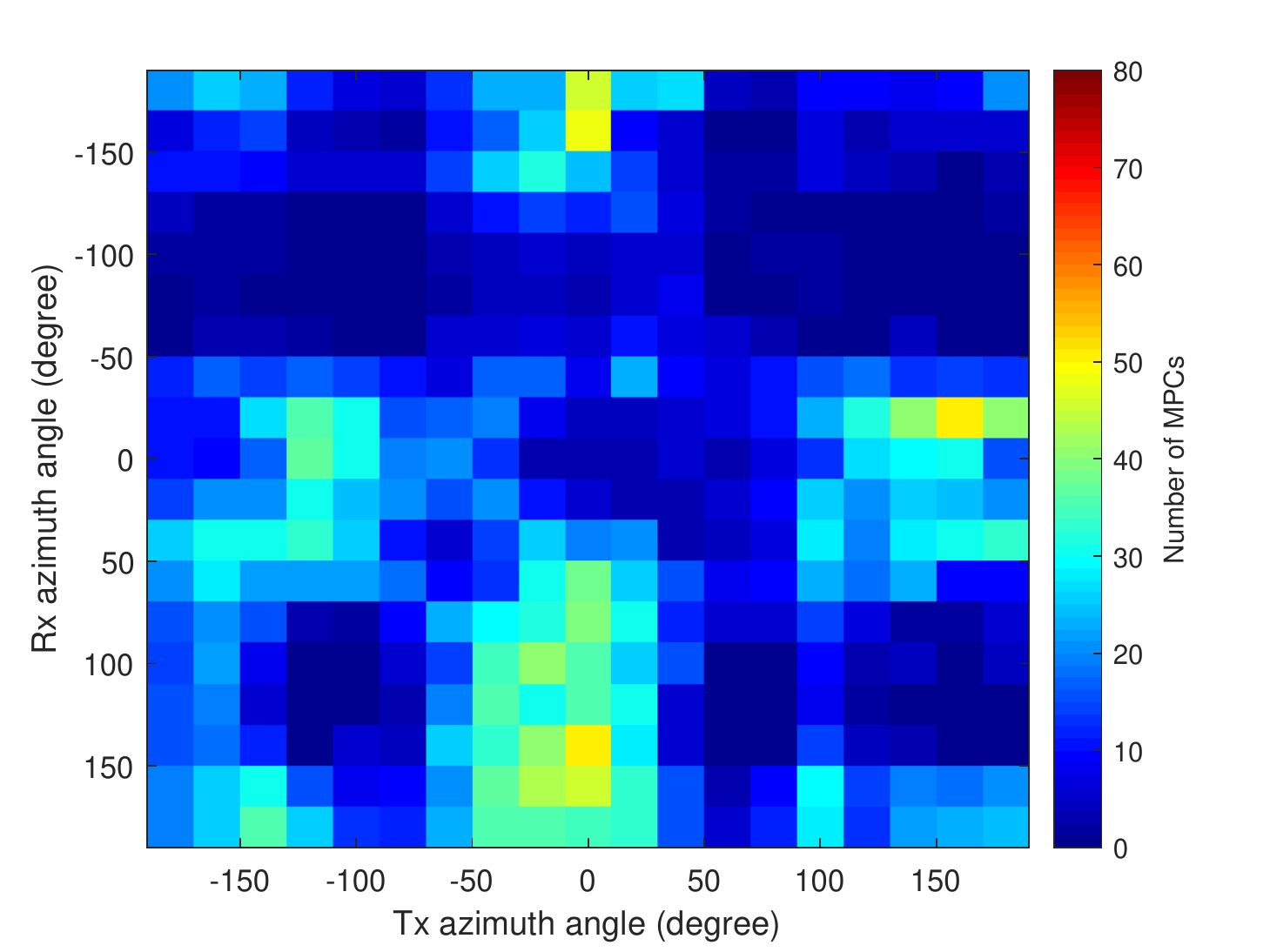} 
	\caption{}
    \end{subfigure}			
	\begin{subfigure}{0.66\columnwidth}
	\centering
    \includegraphics[width=\columnwidth]{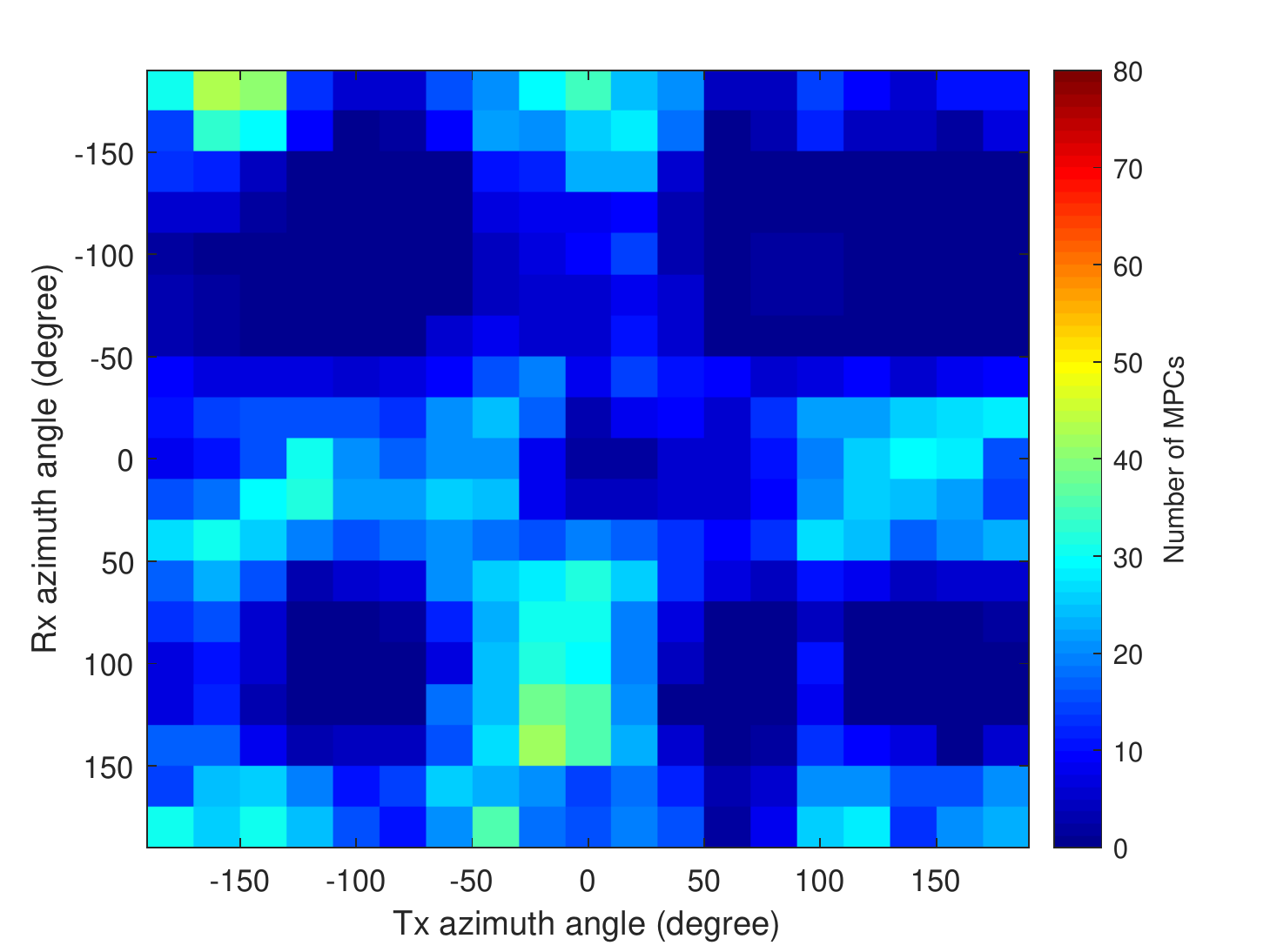}
	\caption{}
    \end{subfigure}
    \caption{Outdoor received power at different Tx and Rx azimuth angle for a Tx-Rx separation of (a) 5.3~m, (b) 9.6~m, (c) 13.9~m. Number of MPCs at different Tx and Rx azimuth angle for a Tx-Rx separation of (d) 5.3~m, (e) 9.6~m, (f) 13.9~m. The angular resolution at the Tx and the Rx is $20$ degrees.}
    \label{out_out_result}
    \vspace{-3mm}
\end{figure*}

\begin{figure*}[h!]
\centering
	\begin{subfigure}	{0.66\columnwidth}	
	\centering
	\includegraphics[width=\columnwidth]{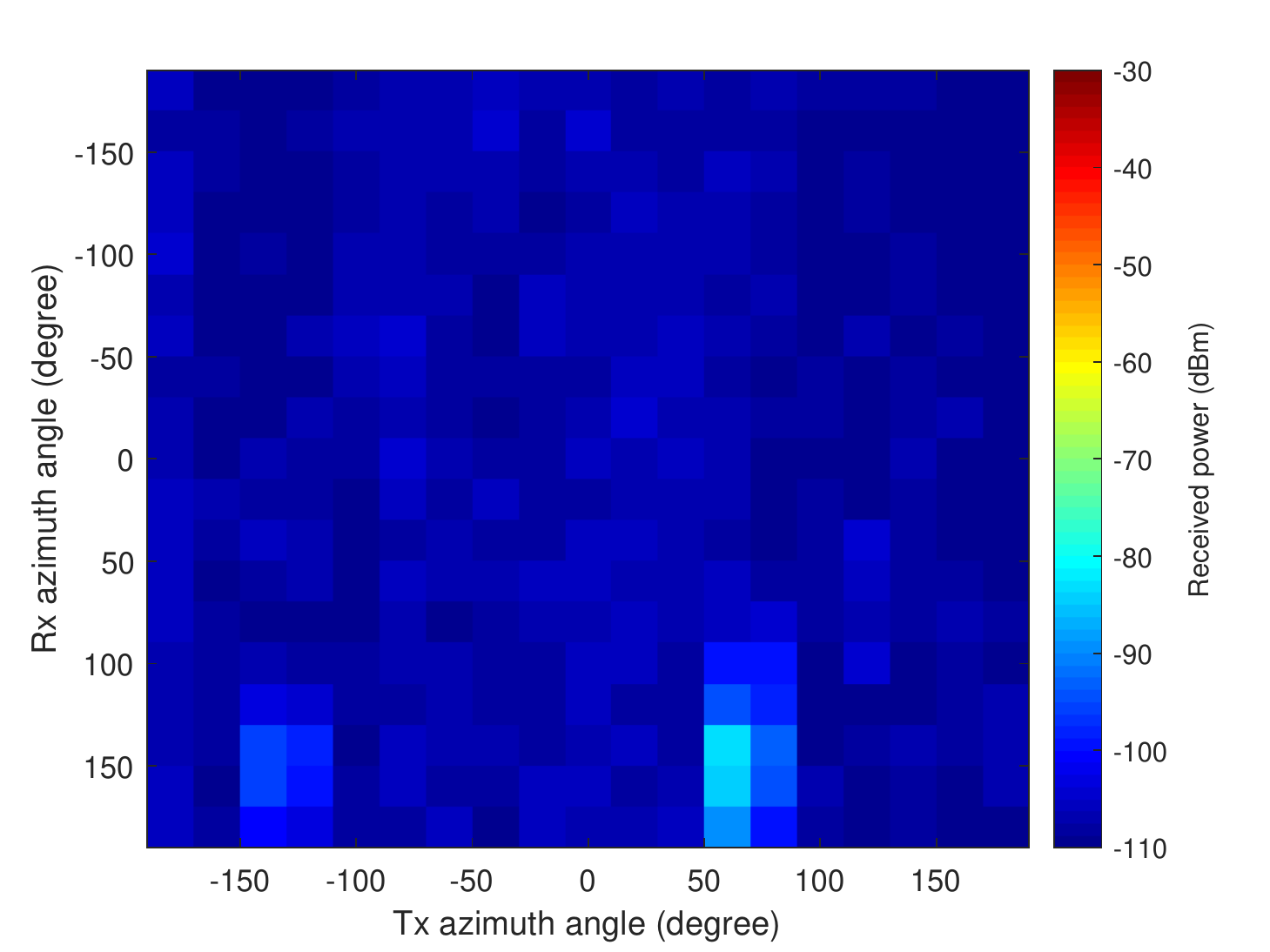} 
	\caption{}
    \end{subfigure}			
	\begin{subfigure}	{0.66\columnwidth}	
	\centering
    \includegraphics[width=\columnwidth]{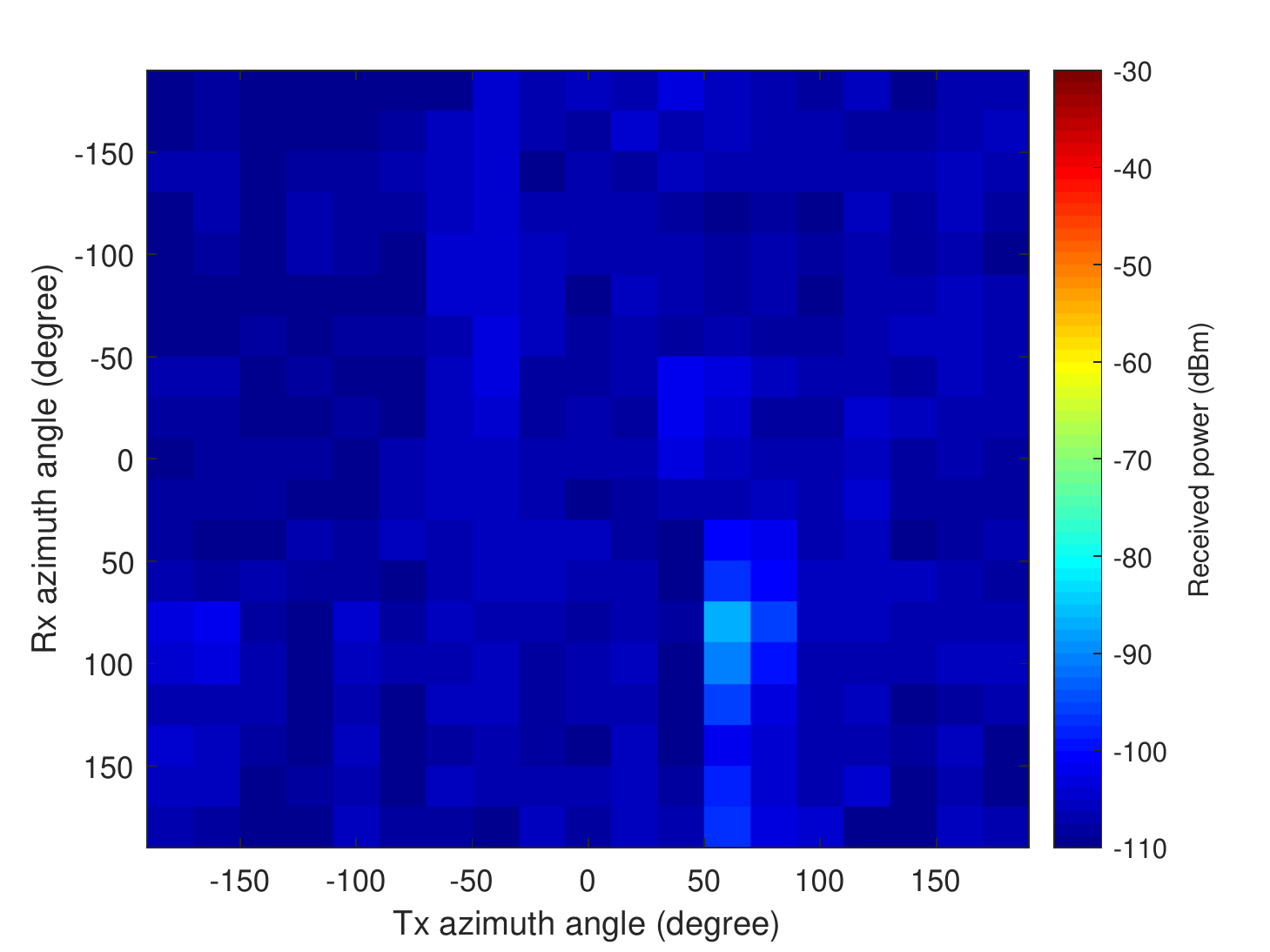}
	\caption{}
    \end{subfigure}
    \begin{subfigure}	{0.66\columnwidth}	
	\centering
	\includegraphics[width=\columnwidth]{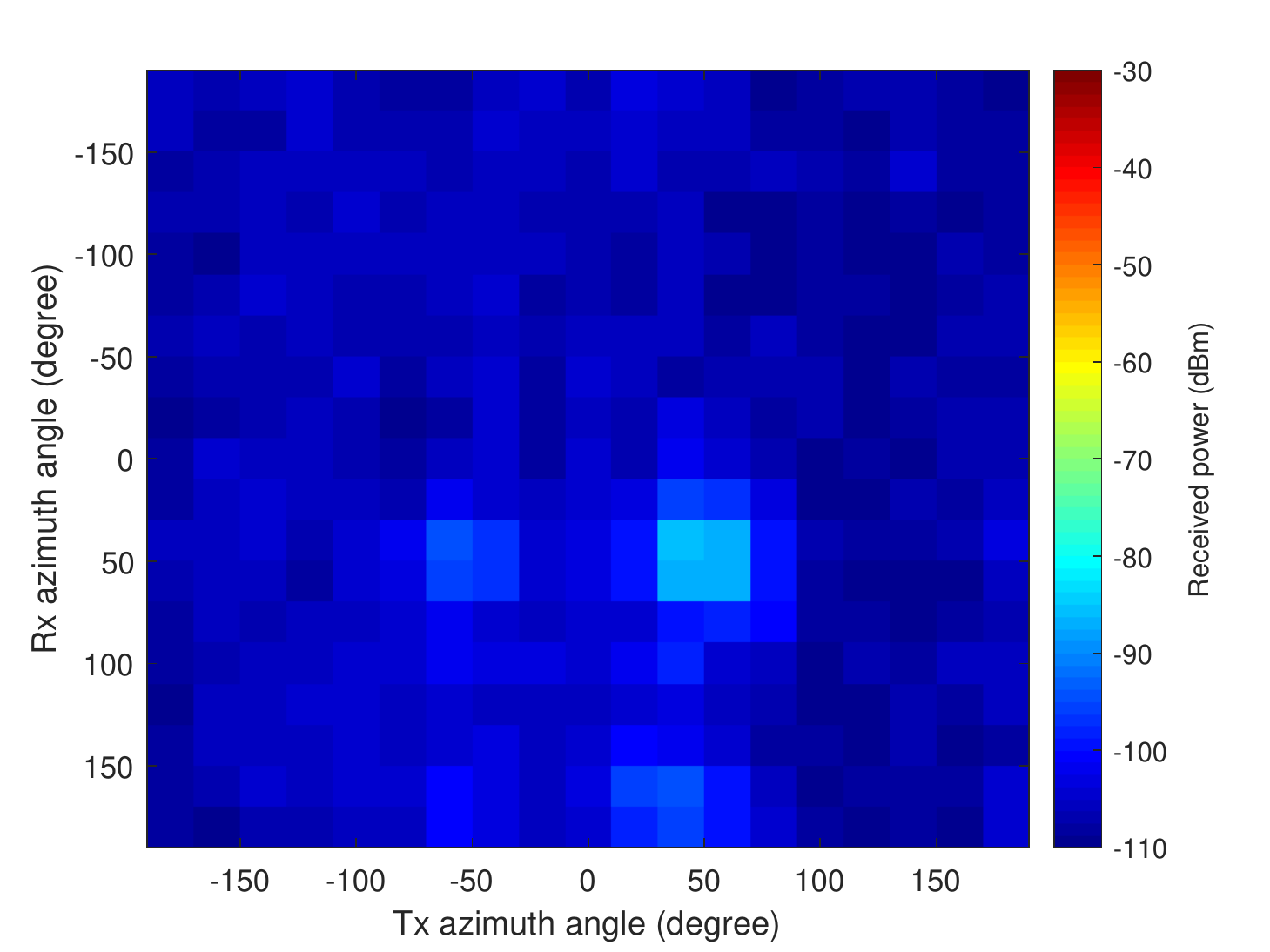} 
	\caption{}
    \end{subfigure}			
	\begin{subfigure}	{0.66\columnwidth}	
	\centering
    \includegraphics[width=\columnwidth]{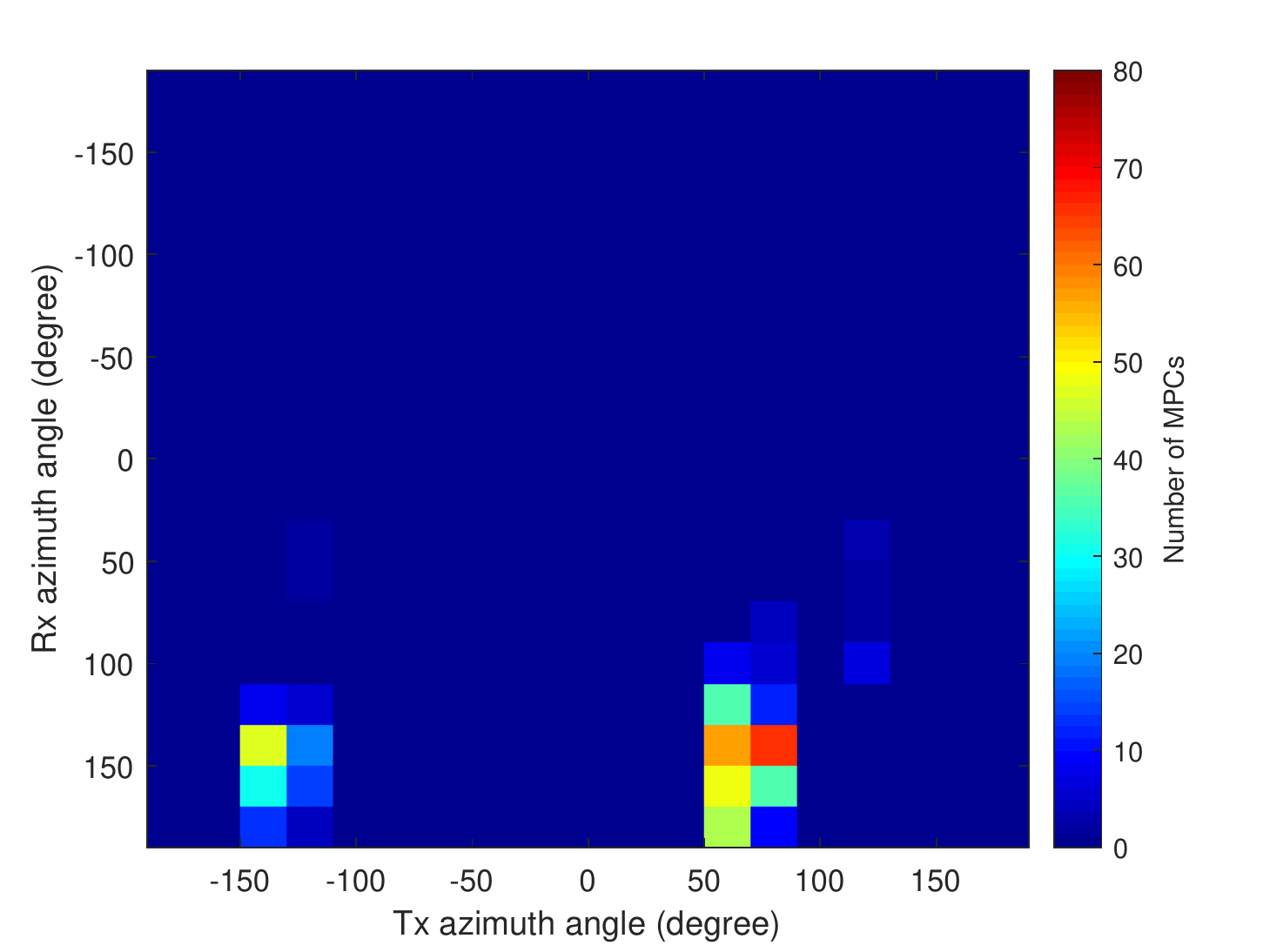}
	\caption{}
    \end{subfigure}	
    \begin{subfigure}{0.66\columnwidth}	
	\centering
	\includegraphics[width=\columnwidth]{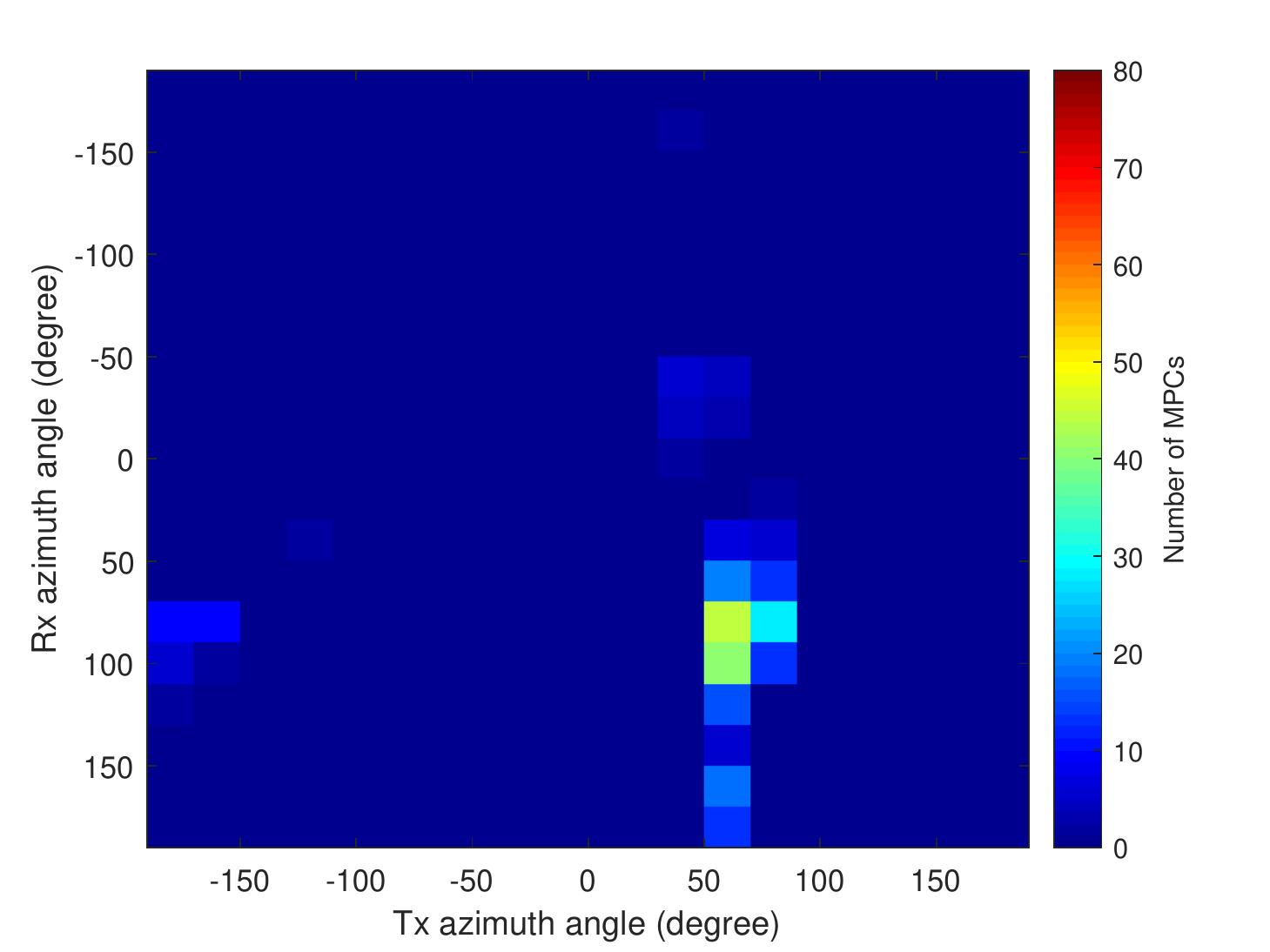} 
	\caption{}
    \end{subfigure}			
	\begin{subfigure}	{0.66\columnwidth}	
	\centering
    \includegraphics[width=\columnwidth]{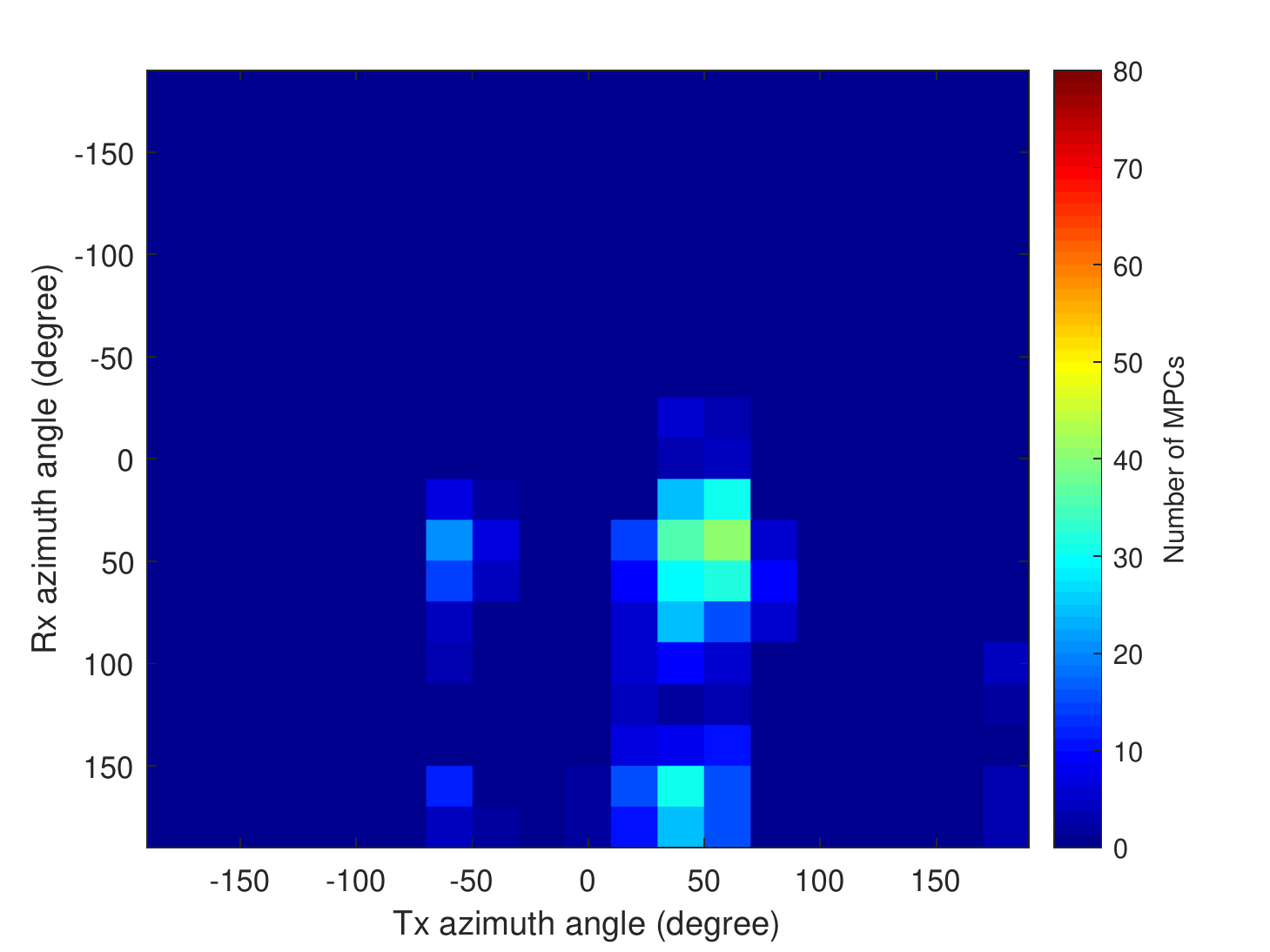}
	\caption{}
    \end{subfigure}
    \caption{Indoor-to-outdoor received power at different Tx and Rx azimuth angle for (a) Rx position 1, (b) Rx position 2, (c) Rx position 3. Number of MPCs at different Tx and Rx azimuth angle for a Tx-Rx separation of (d) Rx position 1, (e) Rx position 2, (f) Rx position 3. The angular resolution at the Tx and the Rx is $20$ degrees.}
    \label{in_out_result}
    \vspace{-3mm}
\end{figure*}

For the outdoor measurement, the number of MPCs in Fig.~\ref{out_out_result} is more stationary compared to indoor environment when the Tx and the Rx separation increases because only certain objects (e.g. wall, ground) could generate MPCs. The received power in the LOS MPC with the azimuth AoA and AoD of 0 degree dominates all the other MPCs. Fewer reflections are observed compared to indoor scenario but some reflected power captured by the receiver is still close to the received power of the LOS MPC.

For indoor to outdoor measurement result shown in Fig.~\ref{in_out_result}, the number of MPCs together with the received power overall have a sharp decrease when compared to indoor propagation and outdoor propagation due to higher penetration loss. Only small portion of the transmitted power is observed in the blocked LOS direction (around the color grid with the highest power). The peak received power suffers more than 90~dB path loss and is approximately 50~dB lower than the peak received power in indoor and outdoor LOS propagation without blockage. Propagation from indoor to outdoor is quite difficult and requires both AoA and AoD aligned to the blocked LOS direction for the received signal to be detectable.

\section{Conclusion}

In this work, we conducted measurements at Johnston Regional Airport to analyze the propagation of 28~GHz mmWave signals. We compared the measurement results in indoor, outdoor, and indoor-to-outdoor scenarios with the the theoretical propagation characteristics. The study showed that mmWave outdoor propagation had a higher FSPL compared to indoor scenario. Indoor environment had rich scatterings and a wider signal coverage while received power in the outdoor airport environment was only dominated by a few rays. Moreover, mmWave had high FSPL and penetration loss, and hence it would be challenging to accomplish indoor-to-outdoor communication: the indoor-to-outdoor propagation therefore may need to be  highly directional to recover the penetration loss through directionality gain. For both indoor and outdoor propagation, there were still a considerable number of reflected MPCs that had comparable received powers to the LOS MPC's received power, which may allow a feasible way to achieve mmWave NLOS communications via the reflective objects in the channel environment.

\bibliographystyle{IEEEtran}
\bibliography{IEEEabrv,reference}
\end{document}